\UseRawInputEncoding
\documentclass{osa-article}
\journal{osac}

\usepackage[title]{appendix}
\usepackage{etoolbox}
\usepackage{epstopdf}
\apptocmd{\appendices}{\apptocmd{\thesection}{: }{}{}}{}{}
\usepackage{graphicx}
\articletype{Research Article}

\usepackage{lineno}
\linenumbers

\begin{document}
\nolinenumbers
\title{Coupling to longitudinal modes in spherical thin shells illuminated by submillimeter wave Gaussian beam: applications to corneal sensing}

\author{Faezeh Zarrinkhat,\authormark{1,2,*} Joel Lamberg,\authormark{2} Aleksi Tamminen,\authormark{2} Mariangela Baggio,\authormark{2} Juha Ala-Laurinaho,\authormark{2} Elsayed E. M. Khaled,\authormark{3,4} Juan Rius,\authormark{1} Jordi Romeu,\authormark{1} and Zachary Taylor,\authormark{2}}

\address{\authormark{1}CommSensLab, Technical University of Catalonia/UPC, Barcelona 08034, Spain\\
\authormark{2}Department of Electronics and Nanoengineering, Millilab, Aalto University, Espoo 02150, Finland\\
\authormark{3}Department of Electrical Engineering, Assiut University, Assiut 71515, Egypt
\\
\authormark{4}High Institute of Engineering and Technology, Sohag 82524, Egypt}

\email{\authormark{*}faezh.zarrinkhat@upc.edu} 



\begin{abstract}
Coupling to longitudinal modes of thin spherical shells, under Gaussian-beam illumination, was explored with a theoretical method based on Fourier-optics analysis and vector spherical harmonics. The illumination frequency band was fixed between $100$ - $600$ GHz and the outer spherical shell radius of curvature and thickness are $7.5$ mm and $0.5$ mm, respectively. The shell material was either the lossless cornea or an aqueous effective media representing the cornea. Six different beam-target strategies were introduced being potential candidates for maximum coupling. Two dispersion-tuned beam ensembles with strongly frequency-dependent phase center location have been created with a fixed incident beam 1/e radius and radius of curvature called forward strategies. These computations of different alignments were continued with four beam ensembles of frequency-invariant phase center, constructed from fits to experimental data, oriented at four different axial locations with respect to the spherical shell center of curvature, they are called reverse strategies. Coupling efficiency for all strategies was calculated for different targets including PEC sphere, air-core, and PEC core covered by a cornea loss-free layer and cornea. All scattering strategies contrasted to scattering from equivalent planar targets as a reference with maximum coupling. The results show that, under an ideal calibration, forward strategies are a closer approximation to the plane-wave condition. However, target properties influence the coupling efficiency remarkably for instance the observed difference in cornea scattering is vanishingly small as dielectric loss limits walk-off loss. Furthermore, under perturbed calibration, the forward strategies showed less sensitivity.
\end{abstract}


\section{Introduction}
The sub-millimeter wave and THz frequency sensing of the cornea leverages the layered tissue structure for corneal water content and thickness quantification. Changes in corneal tissue water content (CTWC) and corneal central thickness (CCT) are correlated with human eye diseases and disorders. Existing clinical measurement approaches are not sufficiently accurate to early detection of these conditions. The way to enhance the accuracy of measurement methodologies is to maximize the reflection from the cornea which could be done by adjusting beam and cornea alignment.
The goals of the paper are to select different beam target strategies which are potential candidates for maximum coupling and investigate these strategies theoretically for different spherical targets. Also, we validated the possibility of using a metal sphere as a calibrator in an experimental set-up.

Up to now, different methods based on a geometrical optic (GO) and physical optic (PO) are presented in corneal sensing literature for beam target analysis although they include approximations and limitations \cite{quasi}. The GO approach leverages stratified media theory under plane-wave assumptions and the measured data is fit to a planar dielectric model under normal incidence plane-wave illumination. In this method, the cornea which is bounded by air and an optically thick body of water on its anterior (outer surface of the cornea) and posterior (bottom layer) segment, is presented as a lossy thin-film lying atop a lossy half-space. Resolution of the cornea’s lossy longitudinal modes via frequency-domain reflectometry in a band sufficiently low (e.g. $220$ GHz - $330$ GHz) for significant penetration allows simultaneous estimates of CTWC and CCT. However, since the cornea is spherical, an efficient coupling to longitudinal modes nominally requires normal incidence across the interrogated area and thus a converging spherical phase front whose curvature matches the corneal surface curvature \cite{zach2015e, zach2015t, zach2018e, zach2018t}. This requirement was not addressed in the aforementioned methodology. On the other hand, the PO approach enables investigating cornea as a spherical surface under Gaussian beam illumination, but it was too intricate and cumbersome to model cornea as a multilayered spherical structure \cite{quasi}.

Sub-millimeter wave and THz sources emit Gaussian beams with standard antenna/optics combinations that produce diffraction-limited spot sizes. As the average corneal radius of curvature (RoC) is $\sim$ $7.5$ mm and the midband free-space wavelength at $100$ GHz - $600$ GHz is $\sim$ $1$ mm, approximate phase-front matching occurs near beam confocal point where the distinctly non-spherical phase-front curvature is rapidly changing and leads to low longitudinal modes coupling. Moreover, the relatively large bandwidth creates complex, frequency-dependent coupling between the beam and the cornea whose thin spherical shells feature wavelength order RoC and thickness.

 The above-listed constraints raise two interrelated questions about obtaining maximum coupling: ($1$) where should the corneal center of curvature (CoC) be positioned relative to beam geometry? ($2$) Is there a frequency dependence on Gaussian beam parameters that further helps to maximize coupling? More succinctly, given a set of constraints on the beam, how do we maximize coupling to the cornea’s sub-millimeter wave longitudinal modes? It's worthy to say maximum coupling is considered as convergence to plane-wave on planar surface coupling due to the perfect match of plane wave's phase-front with the planar surface.

To best address, the above question, a method based on Fourier optics (FO) and vector spherical harmonics (VSH) is used to computationally explore longitudinal coupling in structures resembling cornea. The VSH is the vector solution of the wave equation in spherical coordinates. In 1908, Gustave Mie was one of the pioneers who used VHS to address the incident and scattered field from a sphere illuminated by a plane wave \cite{Mie}. Later, Davis modified the theory for Gaussian beam illumination \cite{davis}. The early proposed method was quite intricate and cumbersome. In 1993, Khaled et al. \cite{esam93} presented a method using Fourier analysis \cite{goodman} to model Gaussian beam as the angular spectrum of plane waves, and for computing scattering coefficients, they employed the T-matrix method \cite{barber}.

In fact, the T-matrix relates the scattering coefficients of a target illuminated by a plane wave with the scattering coefficients of the target illuminated by a Gaussian beam. To compute the scattering coefficients of layered sphere under the plane wave illumination many approaches were presented \cite{Toon},\cite{bohrn},\cite{pena},\cite{Yang}, here Yang \cite{Yang} algorithm was employed. This classical method will serve as a new application in our paper allowing us to scrutinize the cornea as a multilayered spherical structure. FO allows for a closed-form expression of the steady-state scattering solution (multiple reflections within the corneal layer) while VSH provides a convenient representation of the fields with respect to the target geometry.

Here, we applied a computational method to explore different beam-cornea strategies aiming to maximize coupling between the incident and reflected beam. Section 2 quantifies the problem in terms of Gaussian-beam parameters and constructs six different strategies inspired by optimal alignment heuristics and previously published results. The theory to investigate these strategies is described in Sections 3 and 4. Next, the longitudinal modes coupling of a homogeneous PEC sphere, a lossless cornea spherical shell backed by PEC, and backed by air under the six different illumination strategies, are explored and compared to the equivalent planar structure under plane wave illumination. To analyze the phase coupling behavior results are calibrated with the same size PEC sphere. Then, the methodology is repeated for simulated cornea anatomy. Also, each strategy's sensitivity in presence of error in the calibrated sphere is evaluated.

\section{Background and Gaussian beam analysis}

At this point, we explore the constraints on the illuminated beam parameters addressing the questions in the introduction, whether there is a frequency dependence on Gaussian beam parameters assisting coupling enhancement. Also, to determine which beam-cornea alignment are the most likely optimal strategies to explore.

\subsection{Cornea as a spherical scatterer and spherical cavity}

\begin{figure}[htbp]
\centering
\includegraphics[trim=110 410 710 70, clip,width=0.99\textwidth]{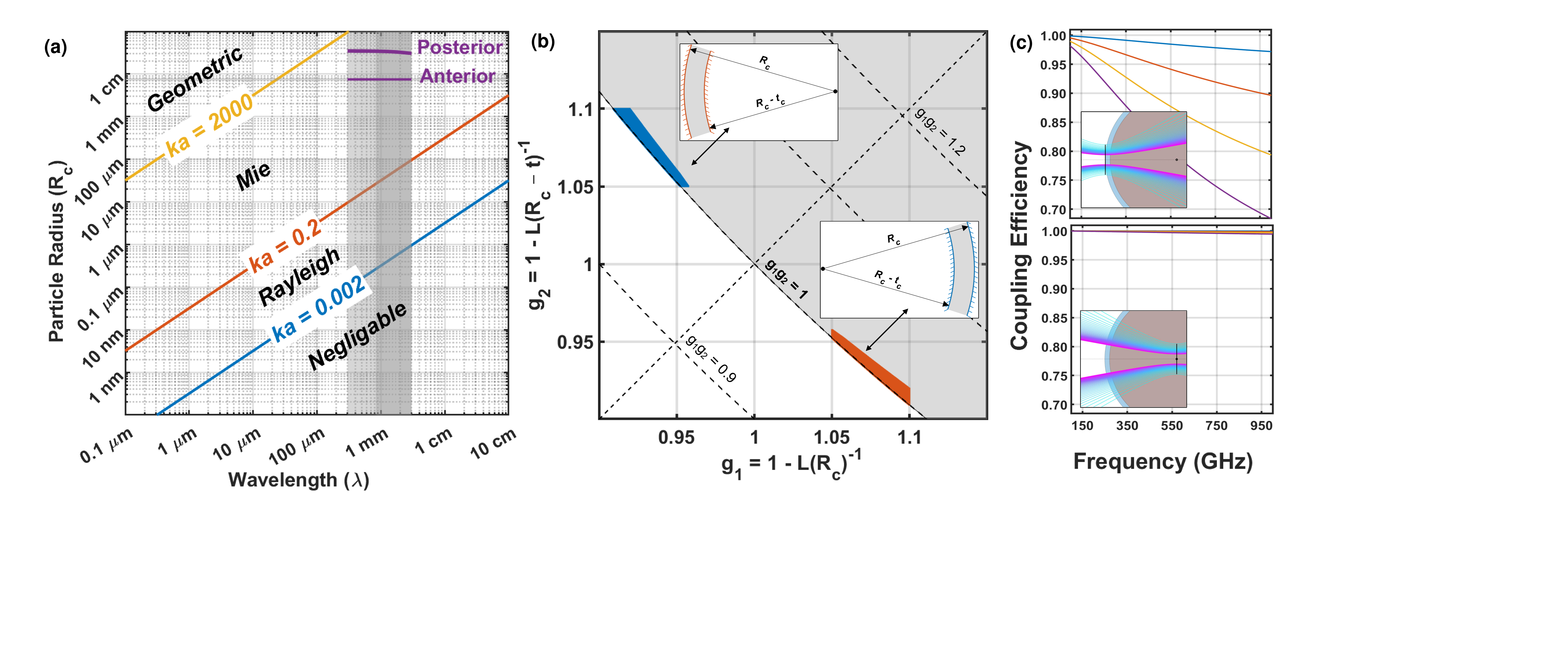}
\caption{(a) Cornea scattering region, (b) stability analysis of the cornea as a spherical cavity, and (c) efficient coupling computed by ABCD matrix}\label{Cavity}
\end{figure}

Interestingly, a cornea can be considered as a scatterer in a wide frequency range. A plot of the size parameter range for cornea in the THz band is shown in Fig. \ref{Cavity}(a) with wavelength along the horizontal axis and particle radius along the vertical axis. The family of oblique lines is parameterized by a fixed size parameter, $ka=2πn/ \lambda$, where $a$, $k$, $\lambda/n$ are respectively, the radius, the wavenumber, and the wavelength in the medium in which the particle resides ($n = 1$ for air refractive index). The fixed size parameter contours of $0.0002$, $0.2$, and $2000$ serve as approximate thresholds between the varying scattering regimes. The Mie scattering is a proper candidate to describe the scattering behavior of particles with size parameters lying in the range $0.2< ka < 2000$.

The vacuum wavelength range $3$ mm - $0.3$ mm corresponding to the $100$ GHz - $1$ THz band is indicated by the gray shaded vertical box. The nominal range for human cornea RoC, $7$ mm - $8$ mm is denoted by the thin purple horizontal shaded rectangle. Its intersection reports the approximate corneal size parameter range subtended by the anterior segment. The size parameter of the posterior segment was estimated by subtracting the nominal CCT range from the anterior segment RoC and computing the wavelength in the complex aqueous corneal medium. Bruggeman's effective media theory \cite{ari} is used to compute the aggregate permittivity of a mixture of $60\%$ collagen and $40\%$ free water. 

These intracorneal size parameters are labeled "anterior segment" and "posterior segment" in Fig. \ref{Cavity}(a). The dispersion arising from the liquid water component is evident in the curvature of the shaded region. At $100$ GHz, both the anterior and posterior segments lie comfortably in the Mie scattering regime. At $1$ THz, the posterior segment is quite close to the approximate geometric scattering limit and, depending on the illumination profile, indicating maybe GO is sufficient to approximate the expected back-scatter.

Moreover, we can treat the cornea as a spherical cavity. It implies certain constraints on the beam. Only particular ranges of the cavity outer radius $R_1$, inner radius $R_2$, and the distance between them $L$, produce stable resonators. An unstable cavity will increase the beam size without limitation, consequently, it will get larger than the cavity size and will be lost completely.

A stability analysis of the cornea for nominal RoCs and thicknesses are displayed in Fig. \ref{Cavity}(b) where $g_1 = 1-L/R_1=0.9333$ and $g_2 = 1-L/R_2=0.9286$. The unstable region $g_1g_2>1$ is indicated by the shaded region and contours of constant $g_1g_2$ by the dotted lines. For clarity, both RoC sign pairs are plotted; ($+R_c, -R_c+t$) and ($-R_c, +R_c-t$). In a region about the corneal apex, the anterior and posterior surfaces are concentric and thus $g_1g_2 = 1$ for all combinations of $R_c$ and $t_c$. The corneal thickness increases towards the periphery which can be described as an anterior segment RoC increase with respect to the posterior segment RoC and thus a divergence from concentricity. This combination is unstable and is represented by the overlap between the $g_1g_2$ curves and shaded unstable region.

Thus, the corneal cavity is susceptible to beam walk-off which reduces the interference between the primary reflection at the air/cornea interface and multiple reflections from paths through the cornea. Two examples of beam walk-off are shown in Fig. \ref{Cavity}(c) using ABCD matrix computations \cite{abcd}. The cornea was modeled via the effective media theory described above but only the real part of the permittivity was applied. Ray transfer matrices were used to compute the parameters of the beam scattered from the anterior segment and then the beams scattered after one, two, and three round trips through the cornea. The plots show the coupling coefficients as a function of frequency between the primary reflection beam and the round trip beams. The superior coupling of the RoC matched beams compared to the incident waist radius beams suggests that the phase front matching will enhance interference and thus aid in the coupling and extraction of longitudinal modes.

\subsection{Gaussian beam analysis}

\begin{figure}[b]
\centering
\includegraphics[trim=0 0 0 0, clip,width=\textwidth]{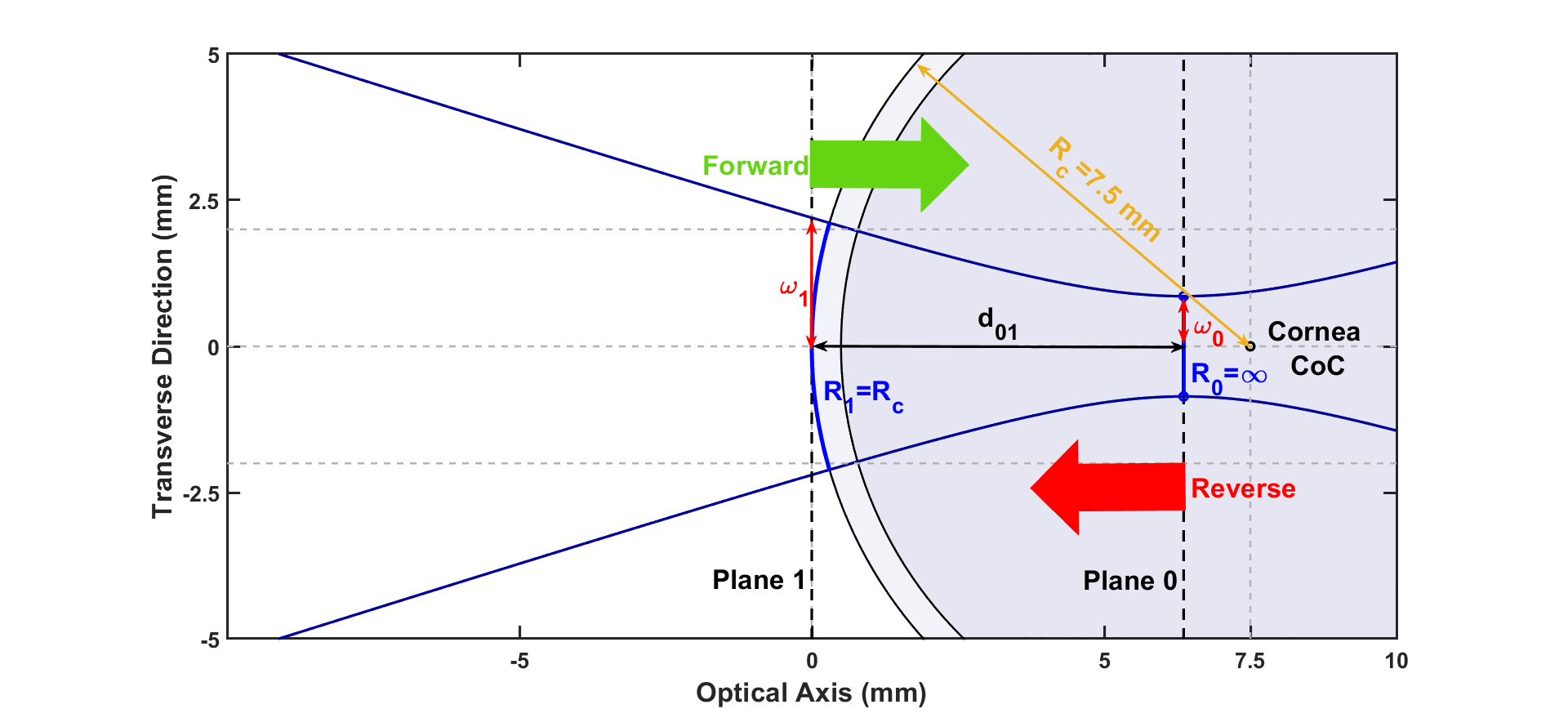}
\caption{Geometry of the problem, propagate beam forwards and backwards
}\label{describe}
\end{figure}

The problem geometry definitions are displayed in Fig. \ref{describe}. A converging beam is an incident, from the left, on the cornea. The incident plane is labeled plane $1$ and the incident beam transverse radius and RoC are denoted $\omega_1$ and $R_1$, respectively. The beam waist is labeled $\omega_0$ and located at plane $0$ where the parameters of plane $0$ and plane $1$ are related via free-space propagation in the absence of the cornea. The physical distance between plane $0$ and plane $1$ is labeled $d_{01}$ and the distance between the corneal CoC and plane $0$ is $z_0$.

Two beam ensemble definitions were considered where the incident beam RoC was fixed such that $R_1= R_c \forall$ frequencies. The first was termed "forward" where $\omega_1$ was set at some constant value and then ray transfer matrices and complex beam parameters were used to compute $d_{01}$ and $\omega_0$ as a function of frequency:

\begin{equation}\label{eq1}
d_{01}(\lambda)=\frac{-R_c(\pi w_1^2)^2}{(\pi w_1^2)^2+(\lambda R_c)^2},
\end{equation}
\begin{equation}\label{eq2}
w_0(\lambda,d_{01})=\sqrt{\frac{((R_c+d_{01})\pi w_1^2)^2+(\lambda d_{01}R_c)^2}{(\pi w_1R_c)^2}}.
\end{equation}

The second approach was termed "reverse" or "varied confocal distance" where $\omega_0$ as a function of wavelength is defined and the standard Gaussian beam equation for axially dependent RoC was used to find $d_{01}$ and then $\omega_1$. Note that $d_{01}$ is the solution to a second order polynomial thus two solutions are possible. These are labeled $d_{01,NF}$ and $d_{01,FF}$ in Eq. (\ref{eqff}) and \ref{eqnf}, respectively where the $d_{01,FF} > d_{01,NF}$ for $Z_c<R_c/2$. The iF subscript stands for $i=N,F$ indicating FF far-field and NF near-field, respectively:

\begin{equation}\label{eqff}
d_{01,iF}(\lambda)=\frac{-R_c\pm\sqrt{R_c^2-4Z_c^2}}{2},
\end{equation}
\begin{equation}\label{eqnf}
\omega_{1,iF}(\lambda,d_{01,iF})=\omega_0\sqrt{1+d_{01,iF}^2/Z_c^2}.
\end{equation}

\begin{figure}[t]
\centering
\includegraphics[trim=150 100 10 10, clip,width=1\textwidth]{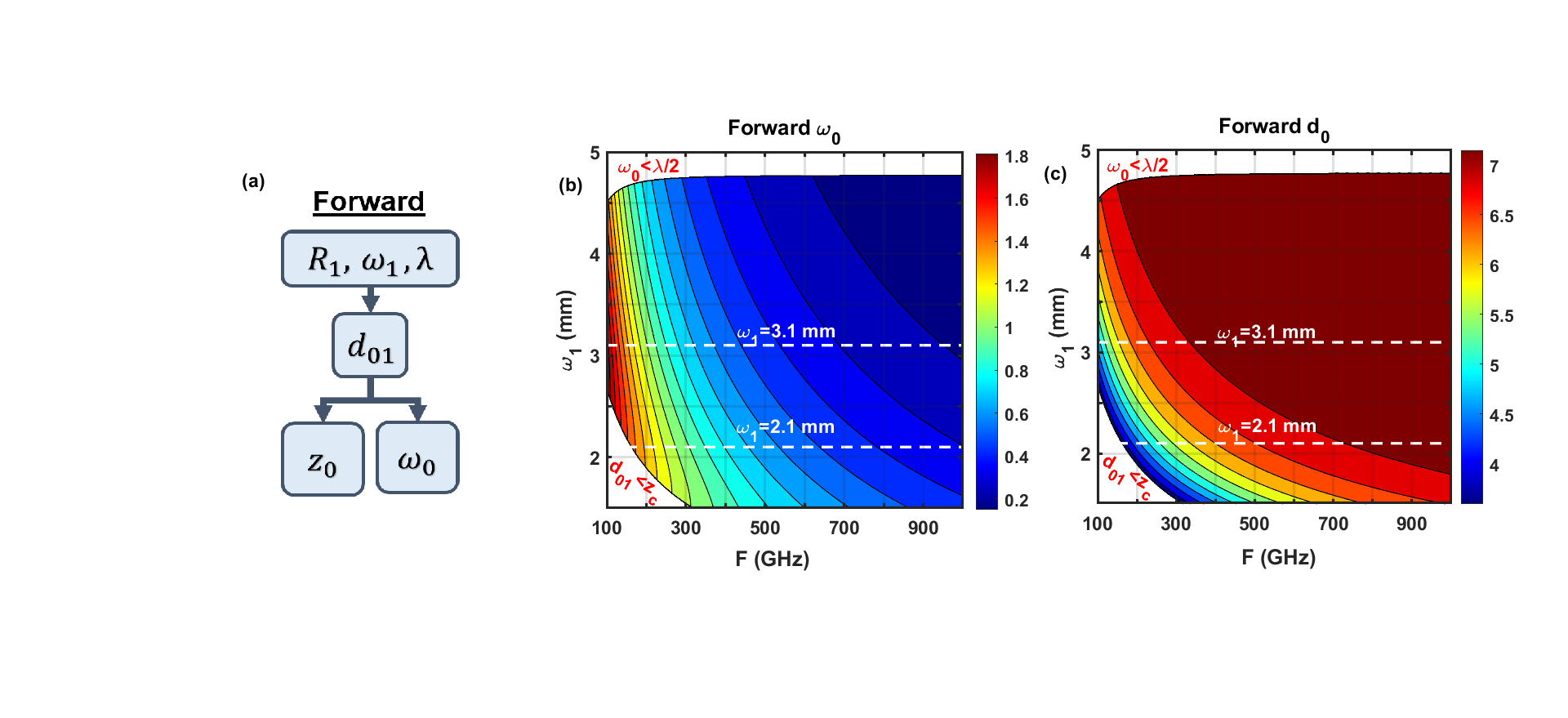}
\includegraphics[trim=150 100 10 100, clip,width=1\textwidth]{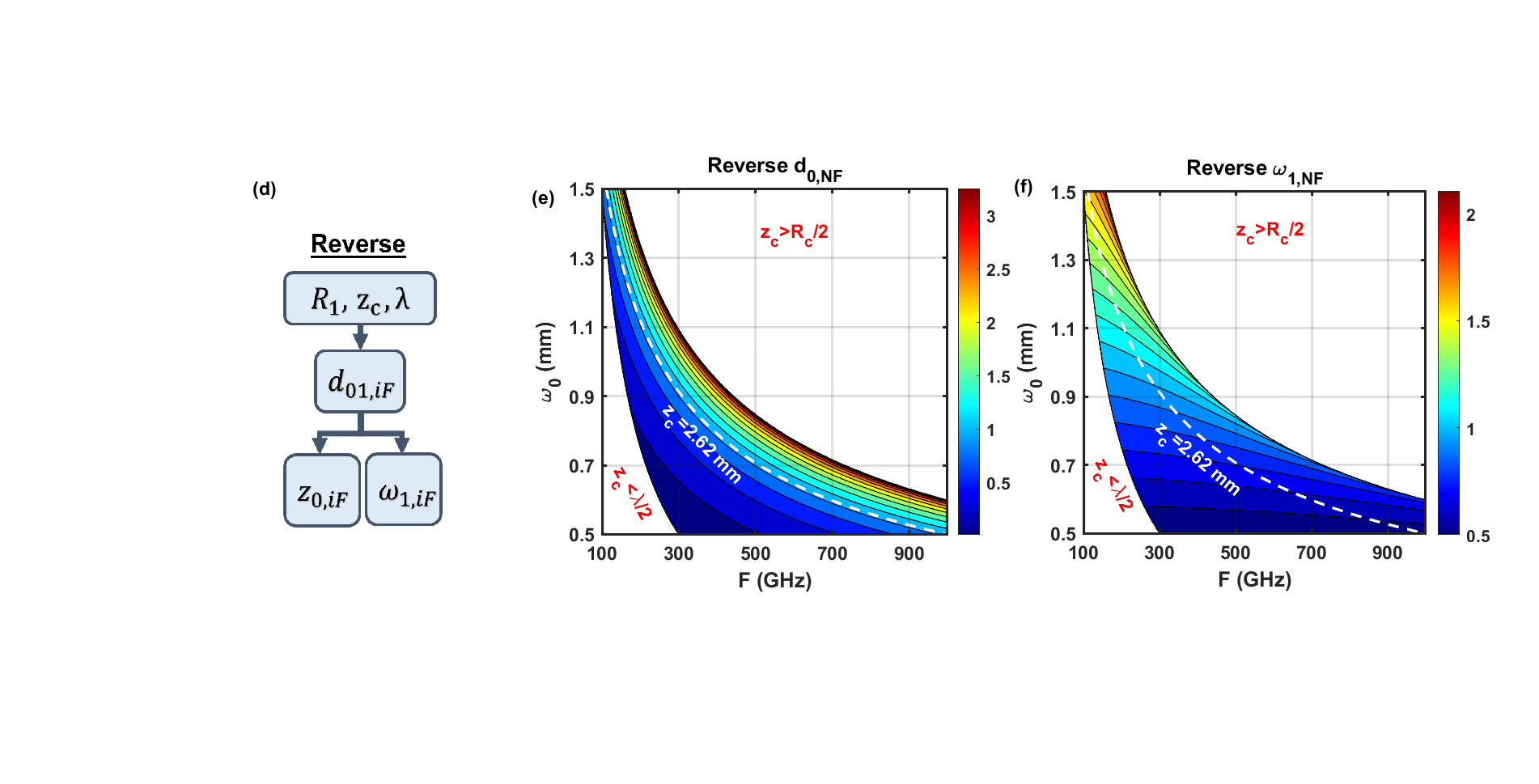}
\caption{Two different beam ensemble definitions (a,b,c) forward and (d,e,f) reverse strategies}\label{limit}
\end{figure}

The $d_{01}$ and $\omega_0$ in the forward direction over a range of $\omega_1$ are reported in Fig. \ref{limit}(a) for $100$ GHz - $1$ THz. In both plots, the ($f$, $\omega_1$) pairs that produce waist radii that violate the paraxial limit $\omega_0< \lambda/2$ are eliminated. Further (f, $\omega_1$) pairs that produce $d_{01} < Z_c$ are denoted. Input beam radii $\omega_1 = 2.1$ mm and $3.1$ mm are denoted by the white dotted lines and are considered in the next section.

Parameters spaces for $d_{01,FF}$ and $\omega_1$ computed in the reverse direction are displayed in Fig. \ref{limit}(b) with the ($f$, $\omega_0$) pairs in violation of the indicated paraxial approximation. The large area in the upper right-hand corner of the plot correspond to ($f$, $\omega_0$) pairs where $Z_c>R_c/2$ and thus complex valued $d_{01}$ has been set aside. Inspection of Eq. (\ref{eqnf}) shows that the contour lines in the $d_{01,NF}$ plot of Fig. \ref{limit} correspond to contours of constant $Z_c$. In other words, if we define $\omega_0$ such that $Z_c$ is invariant to frequency then, as evidenced by Eq. (\ref{eqff}), $d_{01,NF}$ and $d_{01,FF}$ are invariant to frequency and thus the beam RoC evolves equally along the axis for all spectral components.

\subsection{Gaussian beam strategies}

\begin{figure}[t]
\centering
\includegraphics[trim=90 110 90 155, clip,width=\textwidth]{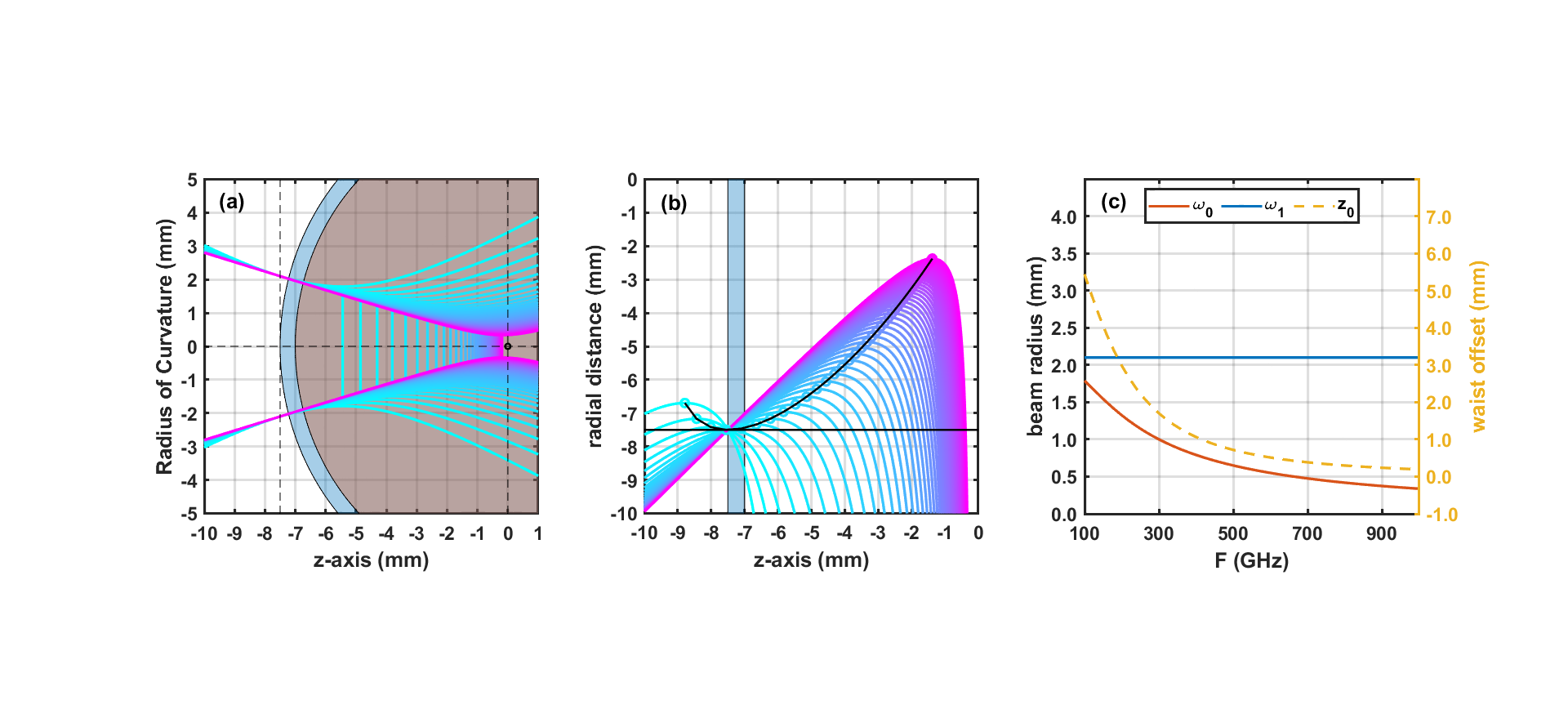}
\vspace*{-7 mm}

\includegraphics[trim=90 110 90 155, clip,width=\textwidth]{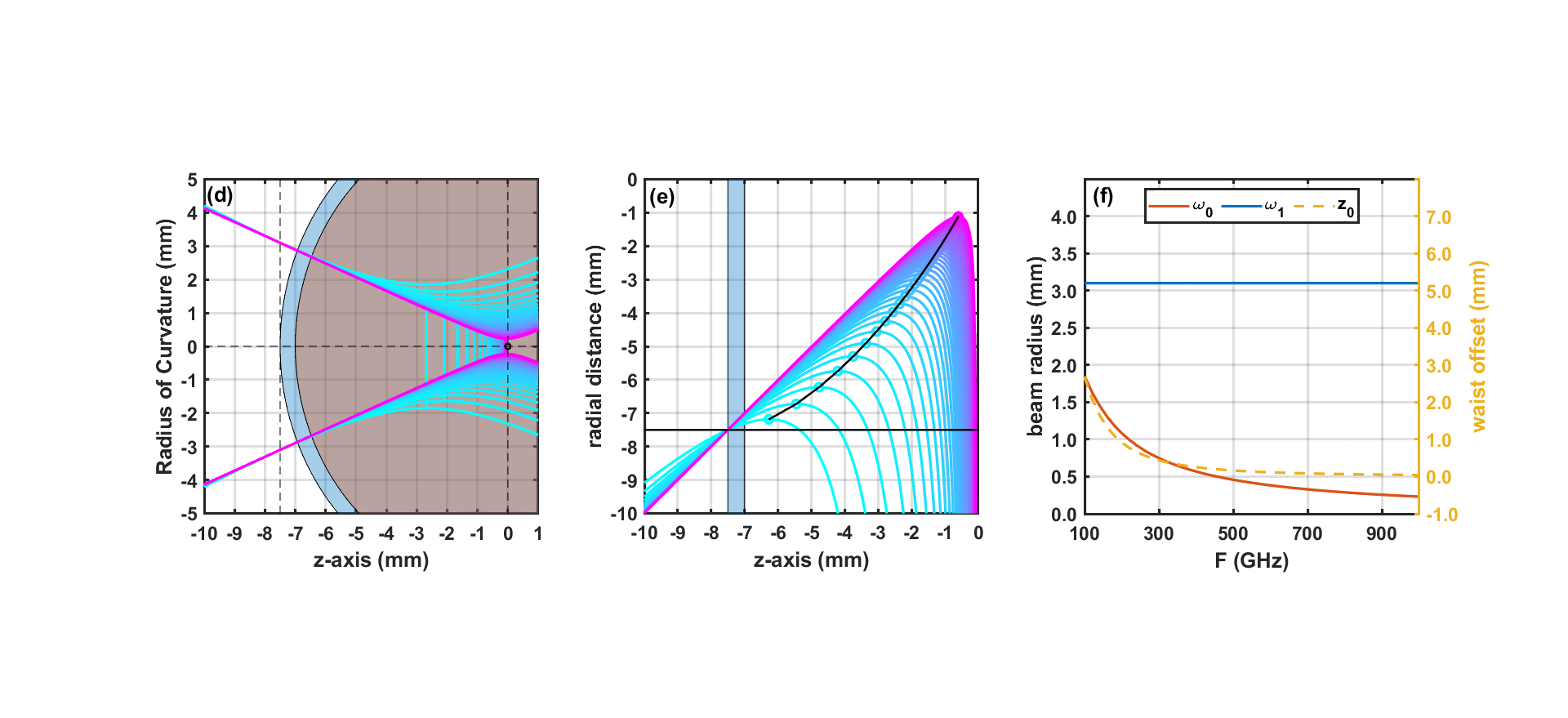}
\caption{Definition and orientation of Gaussian beams and their relationship to corneal geometry, $S_1$ and $S_2$ strategies}\label{s12}
\end{figure}

\begin{figure}[t]
\includegraphics[trim=90 110 90 155, clip,width=\textwidth]{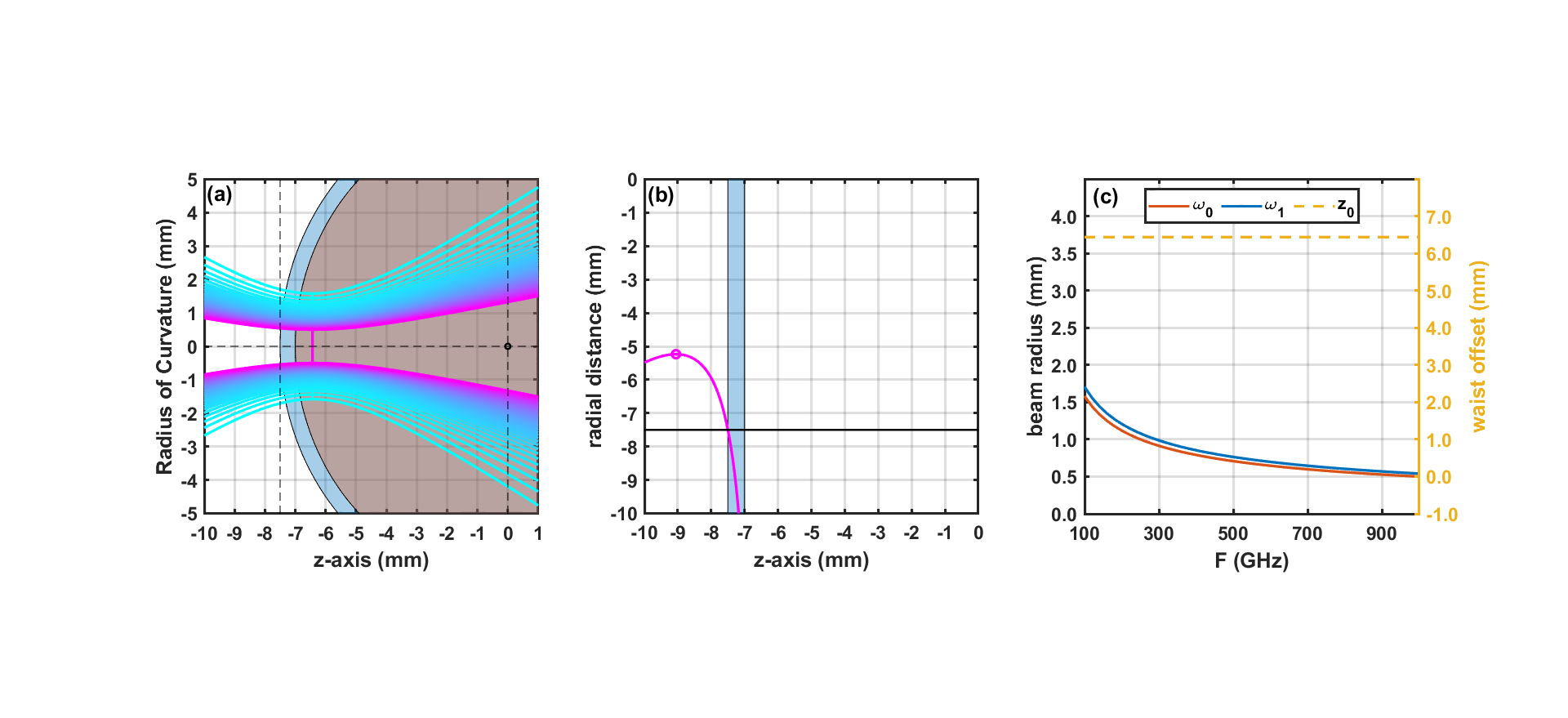}
\vspace*{-7 mm}

\includegraphics[trim=90 110 90 155, clip,width=\textwidth]{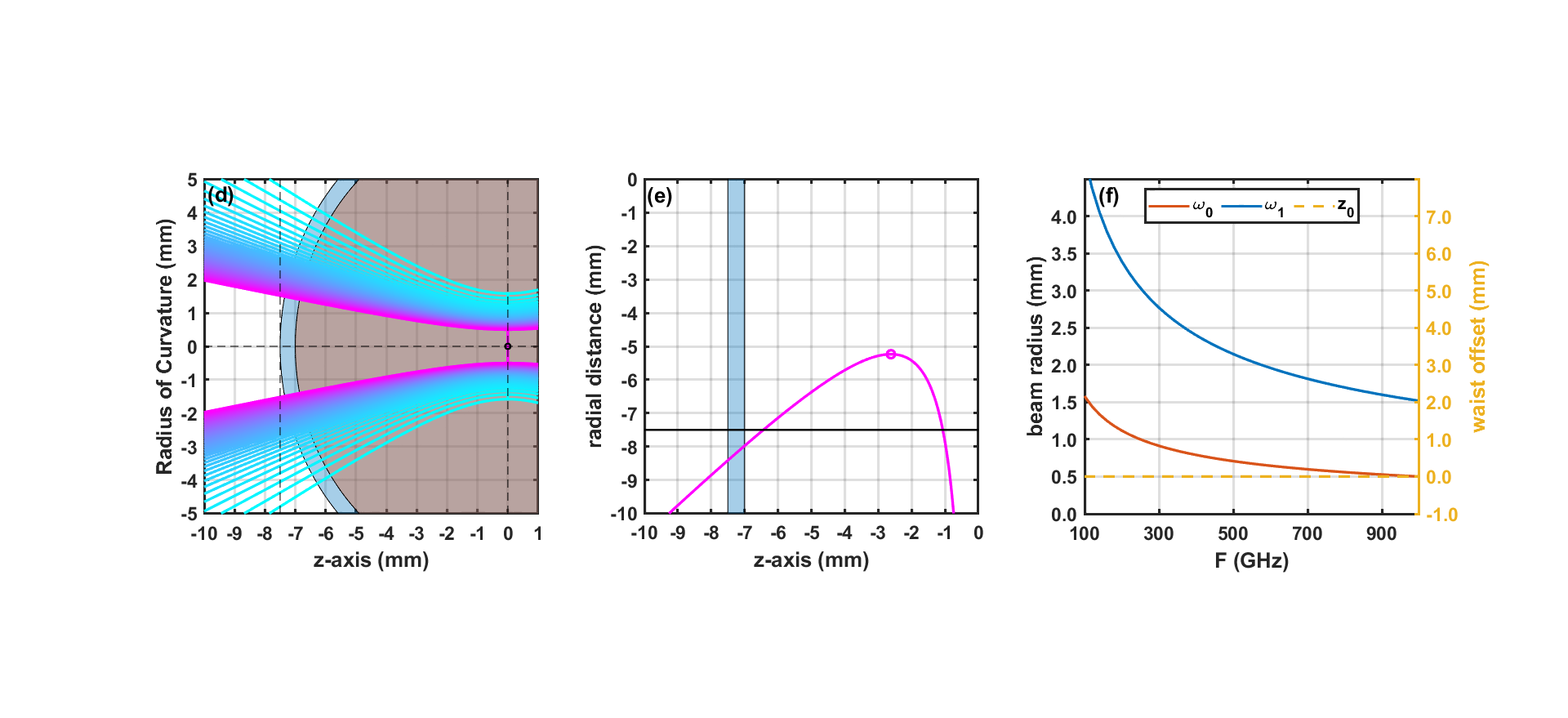}
\caption{Definition and orientation of Gaussian beams and their relationship to corneal geometry, $S_3$ and $S_4$ strategies}\label{s34}
\end{figure}

\begin{figure}[ht]
\includegraphics[trim=90 110 90 155, clip,width=\textwidth]{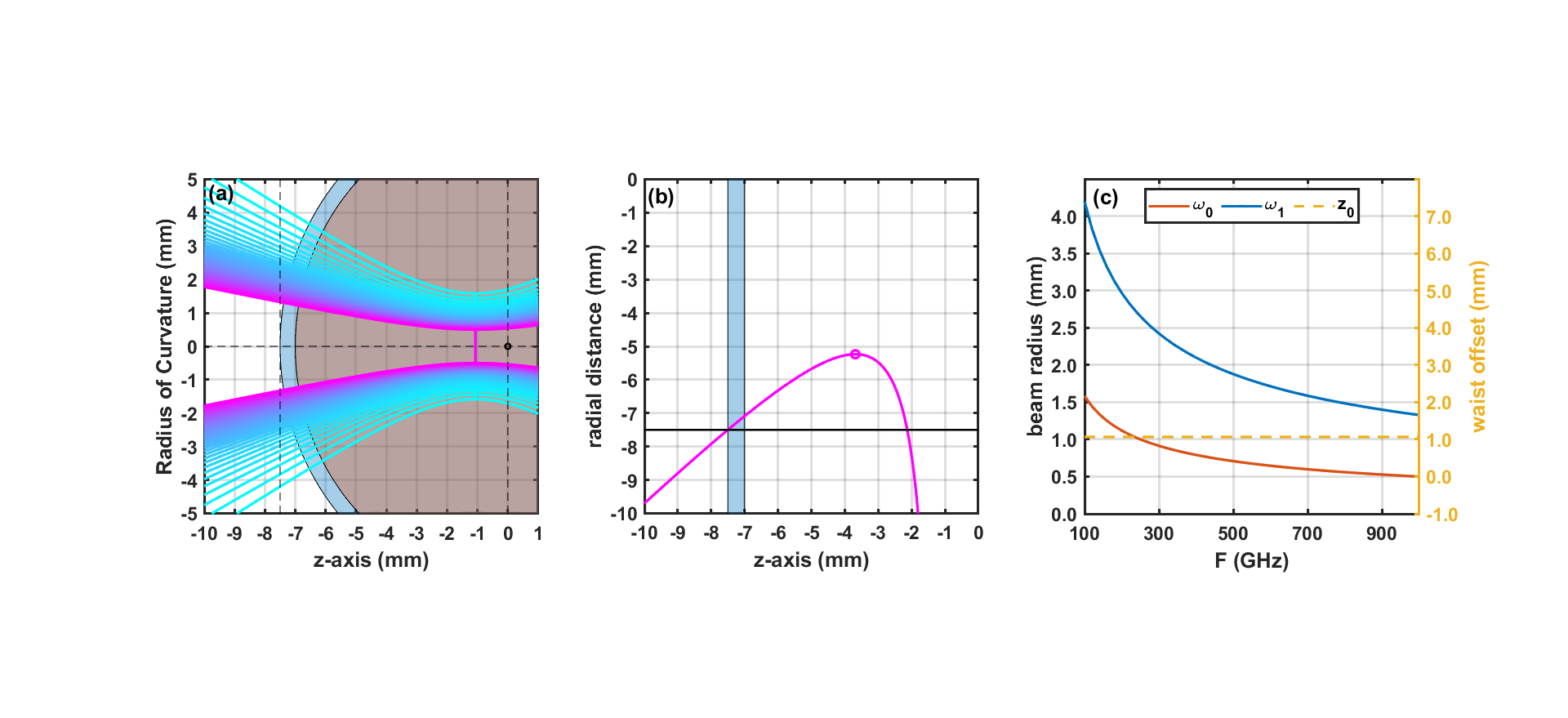}
\vspace*{-7 mm}

\includegraphics[trim=90 110 90 155, clip,width=\textwidth]{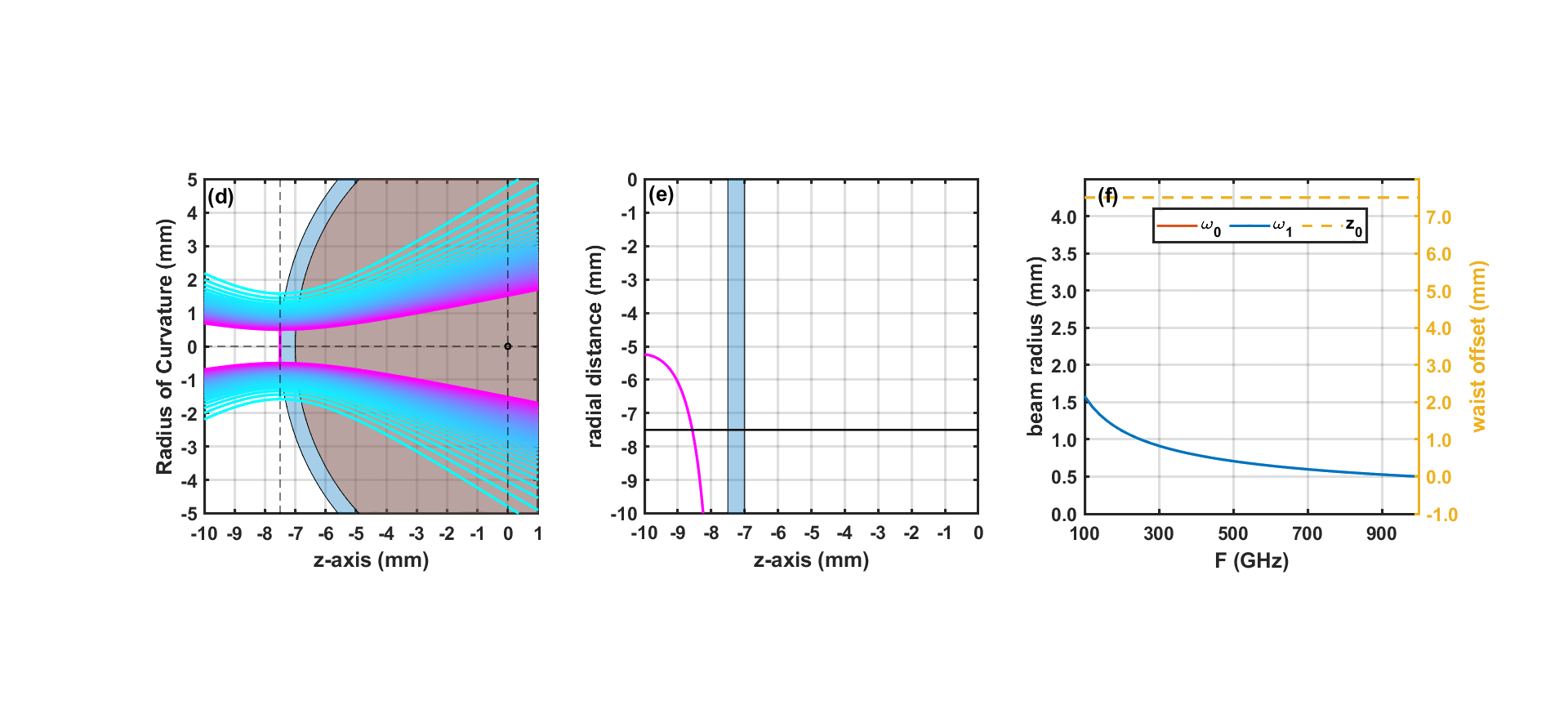}
\caption{Definition and orientation of Gaussian beams and their relationship to corneal geometry, $S_5$ and $S_6$ strategies.}\label{s56}
\end{figure}

Eventually, according to previous discussions, the final candidate strategies are displayed in Fig. \ref{s12}-\ref{s56} where the left column shows the evolution of the beam radius and beam waist location with respect to the corneal geometry and the middle column shows the evolution of the beam RoC superimposed on the corneal anterior and poster surface locations. The right column reports the spectral dependence of beam waist $\omega_0$, incident beam radius $\omega_1$, and beam waist location $Z_0$. 

Strategy $1$ ($S_1$) fixes the input beam radius to $\omega_1=2.1$ mm yielding a diameter similar to that of modern ultrasound pachymeter probes \cite{pach}. Additionally, $S_1$ allows analysis of illumination where the low-frequency incidence occurs in the near field region of the beam and the high-frequency incidence in the far-field. This is evident by the black trace denoting the confocal point location (Fig. \ref{s12}) and requires significant dispersion visible by the beam waist locations reported in the left and right columns.

The beams in strategy $2$ ($S_2$) were also defined in the forward direction with $\omega_1 = 3.1$ mm fixed for all frequency. This ensemble locates the incidence location in the far-field region for all beams but is still sufficiently small to avoid $\omega_0< \lambda/2$. The $S_2$ beam waists are slightly smaller than $S_1$ but the dispersion (variation in $Z_0$) is significantly reduced.

The reverse analysis was utilized for strategies $3$ - $6$ ($S_3$ - $S_6$) with the confocal distance fixed at $Z_c = 2.62$ mm. This confocal distance is below the $R_c/2$ threshold and yields physically realizable, paraxial approximation compatible, providing a set of beams for the frequency range $100$ GHz - $1$ THz. 

The $S_4$ places the phase front match in the super confocal (far-field) region of the beam as evidence by the beam overlapping beam RoC plots in Fig. \ref{s34}. The beam waists are collocated and the waist radii are all larger than $\lambda/2$ although they approach the limit at $100$ GHz. The $S_4$ is the same beam ensemble as $S_3$ but places the phase front match in the subconfocal (near field) region of the beam. The beam radius on the cornea is significantly smaller ($\omega_1 \sim \omega_0$) which should reduce phase front mismatch error in the beam periphery but the beam RoC magnitude is rapidly increasing for increasing $z$ instead of decreasing suggesting a substantial mismatch at the posterior surface.

Strategies $5$ and $6$ ($S_5$, $S_6$) were evaluated for comparison to two strategies commonly reported in the literature. The $S_5$ places the beam waists at the corneal CoC. This is the typical arrangement for imaging via a Gaussian beam telescope optical train and produces phase fronts that are slightly larger in RoC than the corneal RoC. The $S_6$ places the beam waist at the corneal apex thus mimicking the common approach reported by many groups using THz time-domain spectroscopy. The beam radius on target is minimized at the cost of a significant RoC mismatch. We can call them "reference" strategies which are iterations of the reverse strategies.
\section{Fourier-Based Analysis}

An FO based approach, for calculating the coupling efficiency of sub-millimeter wave illumination on the cornea, was presented in \cite{eucap} which utilized the methodology described in \cite{esam93}. This method is applicable for any Gaussian beam incident on a spherical surface and is valid for spheres with RoC in order, or larger than the illumination wavelength. The advantage of FO compared to PO and full-wave approaches are the ability to solve directly the steady state scattering of both homogeneous and coated spheres. This ability significantly reduces computation time as many issues such as "ray splitting" at dielectric interfaces and finely discretized surface current densities can be avoided. Thus, it is possible to explore several illumination strategies on the scattering profile of the cornea when it is modeled as a spherical shell encasing a lossy dielectric sphere.

The applied FO approach first decomposes the Gaussian beam to its plane-wave spectrum representation and then expresses these components as VSH \cite{esam93}. The incident Gaussian beam is represented as: 

\begin{equation}\label{eq3}
\begin{array}{l}
\mathbf{E^i}=\sum\limits_{m}\sum\limits_{n}D[a^e\mathbf{M}^1_{e}+a^o\mathbf{M}^1_{o}+b^e\mathbf{N}^1_{e}+b^o\mathbf{N}^1_{o}],
\end{array}
\end{equation} 
where $a^e$, $a^o$, $b^e$, and $b^o$ are the incident field coefficients (addressed in appendix A) and the VSH of the first kind $\mathbf{M}^1$ and $\mathbf{N}^1$ are the vector solution of the wave equation \cite{bohrn}. The parameter $D = \frac{\epsilon_m(2n + 1)(n-m)!}{4n(n+ 1)(n + m)!}$ is a normalization factor and $\epsilon_m= 1$ for $m = 0$ and $\epsilon_m= 2$ for $m>0$. The radial mode number and azimuthal modes are represented by $n$ and $m$, respectively, and vary over the range $m = 0:N_{stop}$, also $n =m:N_{stop}+1$ ($N_{stop}$ is addressed in Appendix B). For any arbitrary beam-target alignment, the scattered field from the spherical targets can be written \cite{esam93} as: 

\begin{equation}\label{eq4}
\begin{array}{l}
\mathbf{E}^{s}=\sum\limits_{m}\sum\limits_{n}D[f^e\mathbf{M}^3_e+f^o\mathbf{M}^3_o+g^e\mathbf{N}^3_e+g^o\mathbf{N}^3_o],
\end{array}
\end{equation}
where $f^e$, $f^o$, $g^e$, and $g^o$ are scattered field expansion coefficients and $\mathbf{M}^3$ and $\mathbf{N}^3$ \cite{bohrn} are the VSH of the third kind. The scattered fields expansion coefficients are functions of the incident beam coefficients and are calculated by the T-matrix method as \cite{esam94}:

\begin{equation}\label{eq5}
\begin{bmatrix} f^e \\ f^o \\ g^e \\ g^o \end{bmatrix}
=-\begin{bmatrix} T11 & 0 & 0 & 0 \\ 0 & T22 & 0 & 0 \\ 0 & 0 & T33 & 0\\ 0 & 0 & 0 & T44 \end{bmatrix}
\begin{bmatrix} a^e \\ a^o \\ b^e \\ b^o \end{bmatrix}, 
\end{equation}
where, $T$ is a diagonal matrix and its elements, for the case of a coated sphere, are the same scattering coefficients for plane-wave illumination of a coated sphere ($a^c_n$ and $b^c_n$) \cite{esam94}:

\begin{equation}\label{eq6}
\begin{aligned}
T_{11}=T_{22}=-a^c_n,\\
T_{33}=T_{44}=-b^c_n.
\end{aligned}
\end{equation}

For the calculation of coefficients $a^c_n$ and $b^c_n$, Khaled in \cite{esam93} and \cite{esam94} utilized the algorithm introduced by Toon and Ackerman \cite{Toon}. In this work, the algorithm employed by Yang \cite{Yang, pena} was applied to the coefficients calculations. The equations for the algorithm is addressed in Appendix B. Electromagnetic analysis for a layered sphere with more than one layer is the advantage of the Yang algorithm.

\section{Physical-Optics Analysis}

The six different strategies were simulated with an in-house developed physical-optics (PO) script to verify the results of the presented Fourier-optics method. The two-way propagation from an emitting screen to the homogenized PEC sphere and back was simulated with the Gaussian-beam parameters introduced in section $2$. In PO, the field outside the radiating aperture was calculated as in \cite{kong1986electromagnetic}, \cite{orfanidis2002electromagnetic}:

\begin{equation}\label{eq4-1}
\begin{aligned}
\mathbf{E}(\mathbf{r})= \oint_S \nabla G (\mathbf{r}-\mathbf{r^\prime}) \times \mathbf{J}_{ms}(\mathbf{r^\prime})dA,
\end{aligned}
\end{equation}
where $\mathbf{r}$ and $\mathbf{r^\prime}$ are the locations at the radiating aperture and at the observation point respectively, $\mathbf{G}$ is the scalar Green’s function, $\mathbf{J}_{ms}$ is the magnetic surface current density, and $dA$ is the differential area element. The integral was applied first from the virtual Gaussian-beam waist ($\mathbf{r}_{w0}\rightarrow\mathbf{r}_{screen}$) to the screen to define the initial distribution. The two pass propagation from screen to sphere and back to the screen was computed ($\mathbf{r}_{sphere}\rightarrow\mathbf{r}_{screen}$, $\mathbf{r}_{screen}\rightarrow\mathbf{r}_{sphere}$). 

The PO simulated electric field was oriented to ensure no shadowing between the surfaces occurred with an edge taper sufficient to limit the spill-over loss. The coupling coefficient from the PO simulation is overlaid with that from the Fourier-optics method, validate it for homogeneous targets. For all strategies, the coupling coefficient with FO method and PO differ by less than $1\%$ across the band. The coupling coefficient from the physical-optics simulation is consistently less than that from the Fourier method which we believe is due to the remaining spill-over loss necessary to avoid shadowing in this geometry.

\section{Results and Discussion}

 The coupling efficiency of the six different strategies addressed in section 2 was compared for different targets and compared with the plane wave reflection from a multilayered planar surface as the maximum coupling efficiency case. It was computed with Fresnel equations and superimposed on the Gaussian-beam equations to evaluate divergence from the plane-wave condition.

The cornea was modeled as a single-layered spherical shell encapsulating a homogeneous, pure-water sphere. The water permittivity was obtained by the double-Debye model \cite{double}, and the shell (cornea) modeled with the effective medium theory via the Bruggeman mixing model \cite{ari}. The corneal shell consisted of $60\ \%$ water and $40\ \%$ collagen with dispersion-free and real permittivity $2.9$. 

To compare different strategies for various objects, coupling efficiency was defined as Eq. (\ref{eq4-2}). It calculated the coupling between the back-reflected field and the incident field in a plane located at $z=-40$ mm. The integration range for $x$ and $y$ starts from $-4\omega_1$ to $4\omega_1$ and $^*$ denotes the complex conjugate.

\begin{equation}\label{eq4-2}
CE=\frac{\int\int\mathbf{E}^{i}.\mathbf{E}^sdxdy}{\int\int\mathbf{E}^i.\mathbf{E}^{i*}dxdy},
\end{equation}
where $E^{i}$ and $E^{s}$ are computed by Eq. (\ref{eq3}) and Eq. (\ref{eq4}). The magnitude of CE is defined such absolute value of coupling efficiency $|CE|$ and phase CE defines as $\arctan(\frac{image(CE)}{real(CE)})\frac{180}{\pi}$.

Normally, the number of modes ($N_{stop}$) used in CE calculations were determined by the introduced equation in appendix B. To achieve a more robust result, each simulation accuracy was checked in terms of convergence of coefficients to a specific value. For each target, $N_{stop}$ were reported in a table. Simulation time increases exponentially by increasing the number of modes. The frequency band $100$-$600$ GHz was chosen regarding the simulation time and stability of the algorithms as well as constraints introduced in section 2.

\subsection{Homogeneous PEC sphere coupling efficiency magnitude}
\begin{table}[h]
\centering
\caption{Number of modes for PEC sphere and coated PEC sphere simulations}\label{t1}
\begin{tabular}{ c | c  c }
\hline
\textbf{Frequency (GHz)} & 100-445 & 445-600 \\ 
\hline
$\mathbf{N_{stop}}$ & 75 & 95 \\ [0.5ex] 
\hline
\end{tabular}
\end{table}

\begin{figure}[t]
\centering
\includegraphics[trim=90 125 90 140, clip,width=\textwidth]{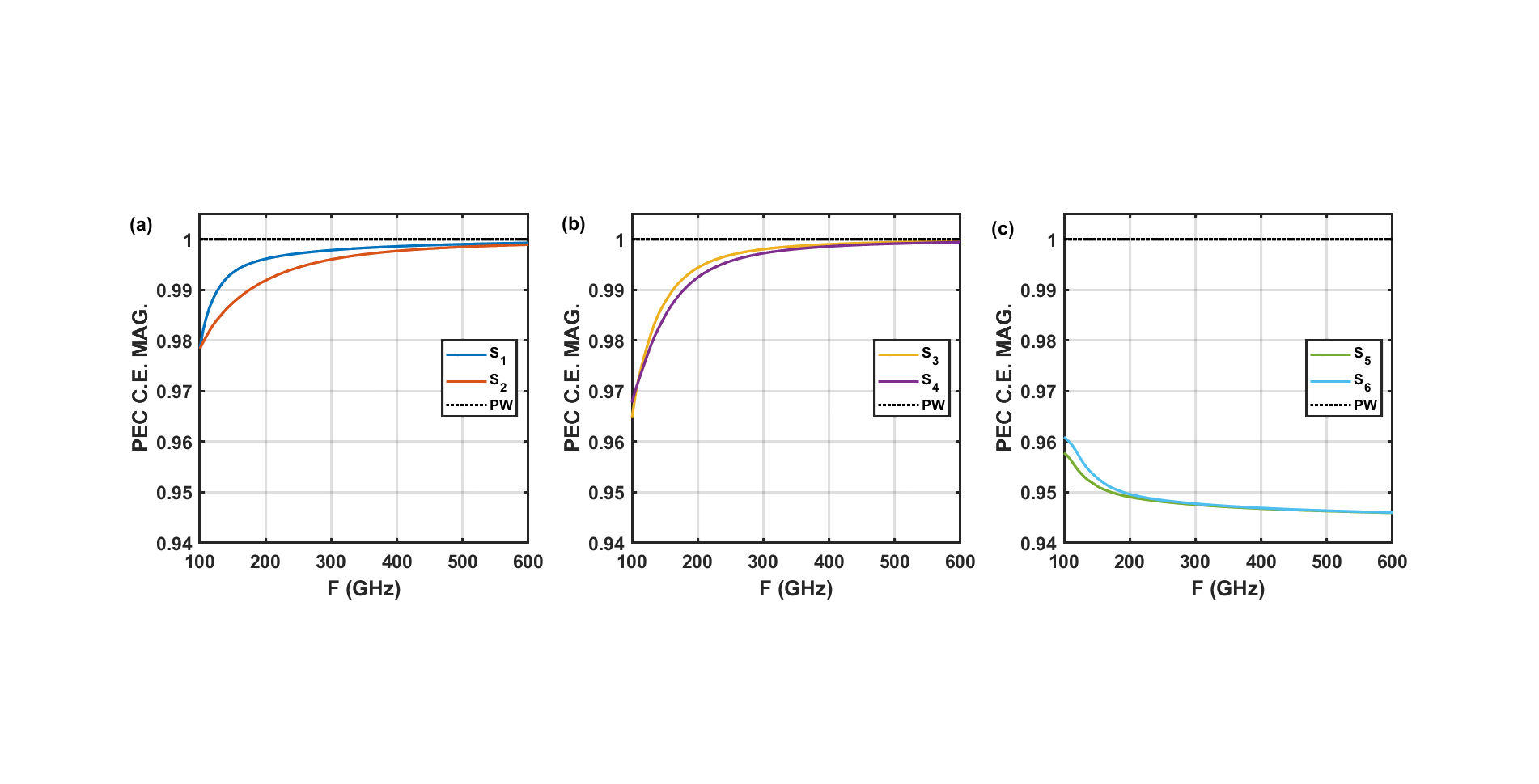}
\caption{ PEC sphere coupling efficiency magnitude comparison is presented for (a) Forward strategies, $S_1$, and $S_2$, (b) reverse strategies, $S_3$, and $S_4$, and (c) reference strategies, $S_5$, and $S_6$. Plane wave reflection from equivalent PEC planar is indicated by black-dotted line and radius of sphere equals $R_c=7.5$ mm.}
\end{figure}
In this subsection, the coupling efficiency of the previously described designs in section 2 for the PEC sphere was calculated using the proposed methodology in section 3. The Gaussian-beam illuminating PEC was assumed to scatter more energy back at the screen than any other material of equal RoC and thus served as a reference/calibration target for the further exploration of PEC coated with the lossless cornea, air-core with lossless cornea shell, and cornea simulations.

To compute the scattering coefficients of a PEC sphere illuminated by a Gaussian beam, it was first necessary to compute the scattering coefficients of the PEC sphere when illuminated by a plane wave according to Eq. (\ref{eq6}). Here, the approach in \cite{pec} and presented in appendix C was applied. The PEC sphere radius was set to $R_c=7.5$ mm and the beam waist radii locations were distributed to match strategies $S_1$ to $S_6$ as described in section 2. The number of modes to reach enough accuracy for the PEC sphere is reported in table \ref{t1}. The simulation time for each strategy was about 17 hours on UPC computational cluster.

 The forward strategies $S_1$ and $S_2$, with dispersion set to achieve frequency wavefront matching, behave similarly and increase from $97.84 \%$ to $99.92 \%$ and from $97.83 \%$ to $99.89 \%$. Both reverse strategies $S_3$ and $S_4$ coupling efficiencies behave almost in the same way and range from $96.46 \%$ to $100 \%$ and $96.77 \%$ to $99.94 \%$ across the band. The reference strategies $S_5$ and $S_6$ behave likewise and unlike the other strategies decrease over the frequency band from $95.71 \%$ to $94.59 \%$ and $96.08 \%$ to $94.59 \%$, respectively.
 
Overall, for homogeneous PEC sphere being a target, $S_1$ comes closest to the plane-wave condition although it behaves closely to $S_2$. Forward strategies reveal higher coupling compared to reverse strategies ($\sim 1.1 \%$) and reverse strategies display more coupling compared to reference strategies ($\sim 0.4-5.5 \%$ across frequency band). 

\subsection{Lossless cornea sitting on a PEC and air core coupling efficiency}
\begin{figure}[ht]
\centering
\includegraphics[trim=90 125 90 145, clip,width=\textwidth]{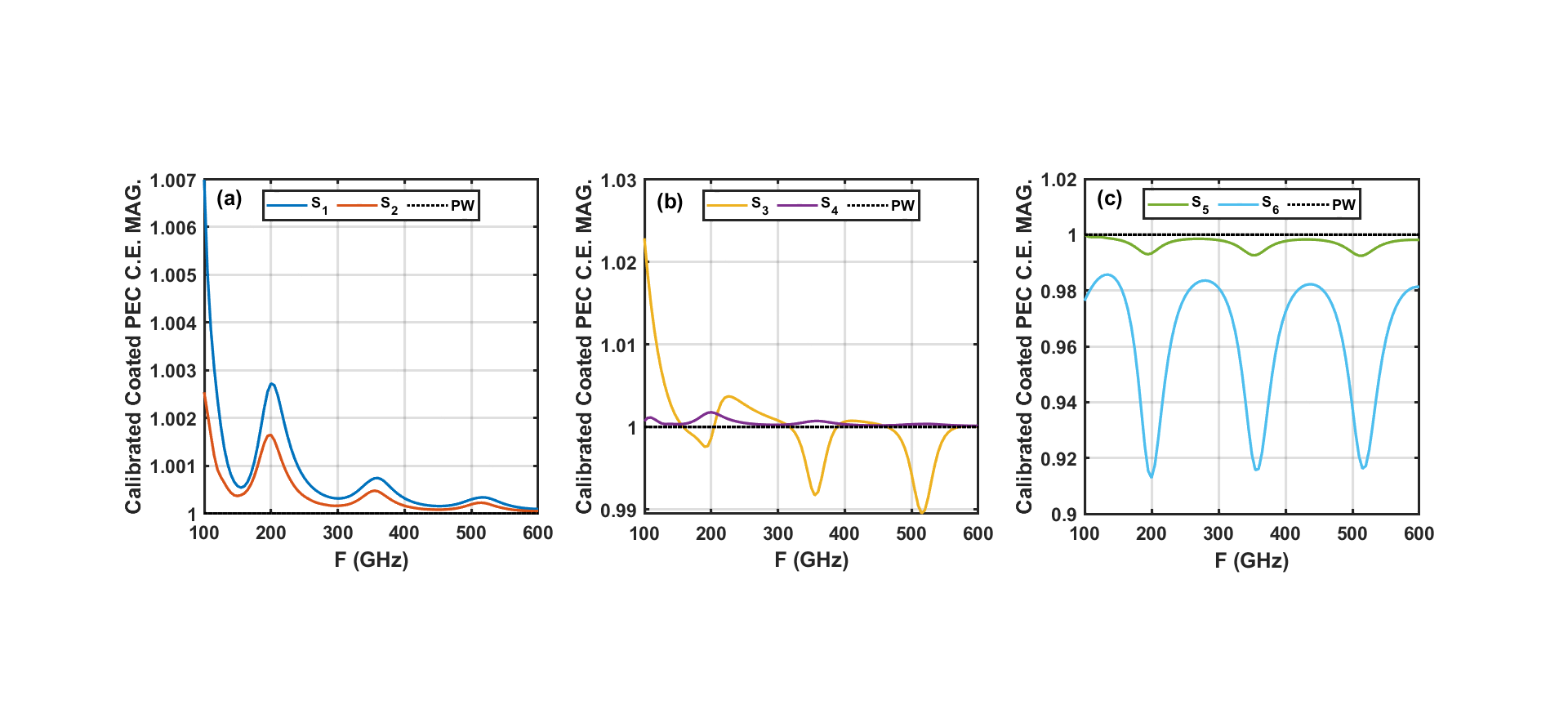}
\vspace*{-6 mm}

\includegraphics[trim=90 125 90 145, clip,width=\textwidth]{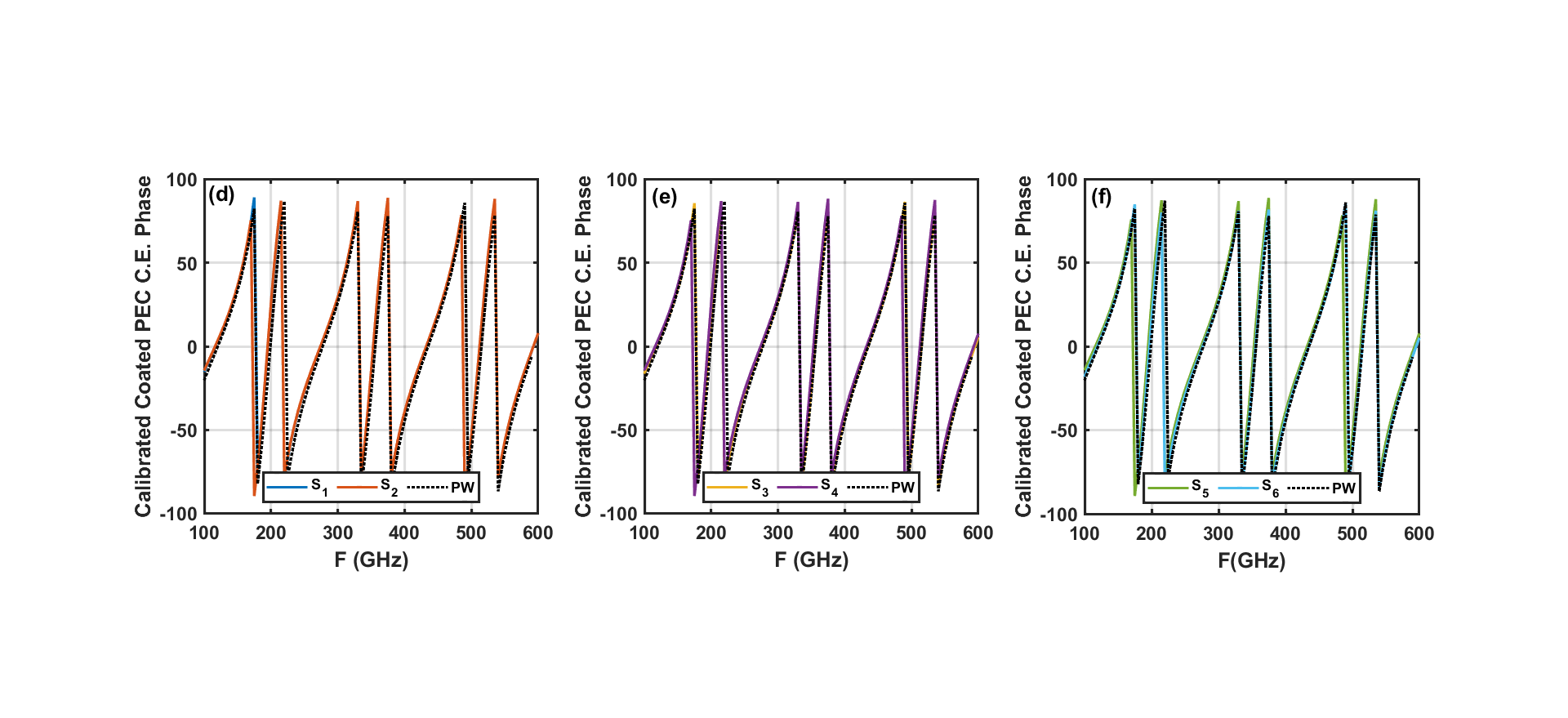}
\caption{Calibrated lossless cornea sitting on a PEC core coupling efficiency magnitude and phase comparison are presented for (a,d) Forward strategies, $S_1$, and $S_2$, (b,e) reverse strategies, $S_3$, and $S_4$, and (c,f) reference strategies, $S_5$, and $S_6$. Plane wave reflection from equivalent layered planar is indicated by black-dotted line. The shell permittivity is set according to real part of cornea permittivity and its thickness is $0.5$ mm sitting on $7$ mm RoC PEC sphere.}\label{Q}
\end{figure}

Lossless dielectric spherical shells backed by PEC were simulated to explore the effect of spectral phase-front variation and mismatch. A target was constructed of $0.5$-mm thick lossless cornea spherical shell encapsulating a PEC sphere of $7$ mm RoC. The layer refractive index was set as the real part permittivity of the cornea computed by effective media theory \cite{ari}. The coupling efficiency was computed with Eq. (\ref{eq4-2}). Then, the calibrated coupling efficiency was computed by normalizing the PEC-backed lossless cornea shell coupling efficiency with the coupling efficiency from a $7.5$ mm PEC sphere (Fig. \ref{Q}). The scattering coefficients of a coated PEC were addressed in appendix C. The number of modes considered to reach stability for coated PEC was similar to PEC sphere modes in table \ref{t1} and simulation time was the same.

This operation mimics experimental calibration routines \cite{quasi} and enables analysis of the absolute phase following deconvolution of the free space path influence on complex coupling angle. The PEC sphere calibrated lossless cornea shell coupling efficiency magnitude and phase are displayed in Fig. \ref{Q} where magnitude and phase of forward strategies $S_1$ and $S_2$, reverse strategies $S_4$ and $S_5$, and reference strategies $S_5$ and $S_6$ are plotted in panels (a), (b) and (c), respectively. All trends are referenced to the equivalent plane-wave condition.

Interestingly, as illustrated in Fig. \ref{Q}, with calibrating all strategies, phase behaviors were almost consistent with plane wave condition, implying a proper phase matching. In fact, they matched perfectly if plane wave phase figure shifts about $10$ GHz toward low frequencies. These results suggest regardless of illumination profile, as long as we calibrate them with the same size RoC PEC sphere, we can get a decent phase match and resolution.
Moreover, from the magnitude aspect, forward strategies provided better coupling between the incident and back-scattered fields than reverse ones, however, $S_4$ featured closer to $S_1$ and $S_2$ (less than $1 \%$ difference). It could be said $S_3$ and $S_5$ were also close enough to plane wave condition (less than $\sim 2.2 \%$ difference) and certainly, $S_6$ was far from proper coupling ($1.5-8.8 \%$ deviation). 

\begin{figure}[ht]
\centering
\includegraphics[trim=100 80 100 90, clip,width=0.75\textwidth]{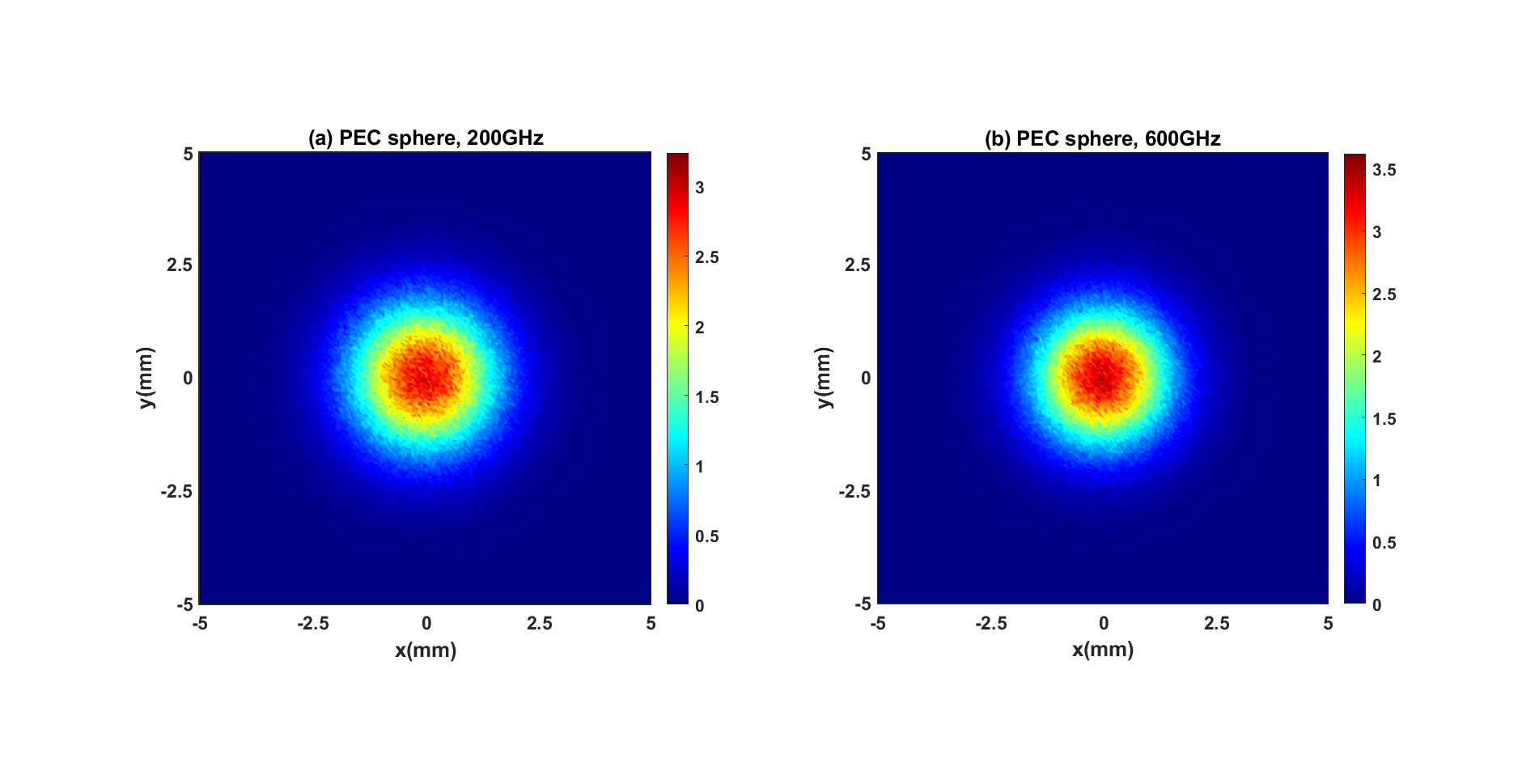}
\vspace*{-1 mm}

\includegraphics[trim=100 80 100 90, clip,width=0.75\textwidth]{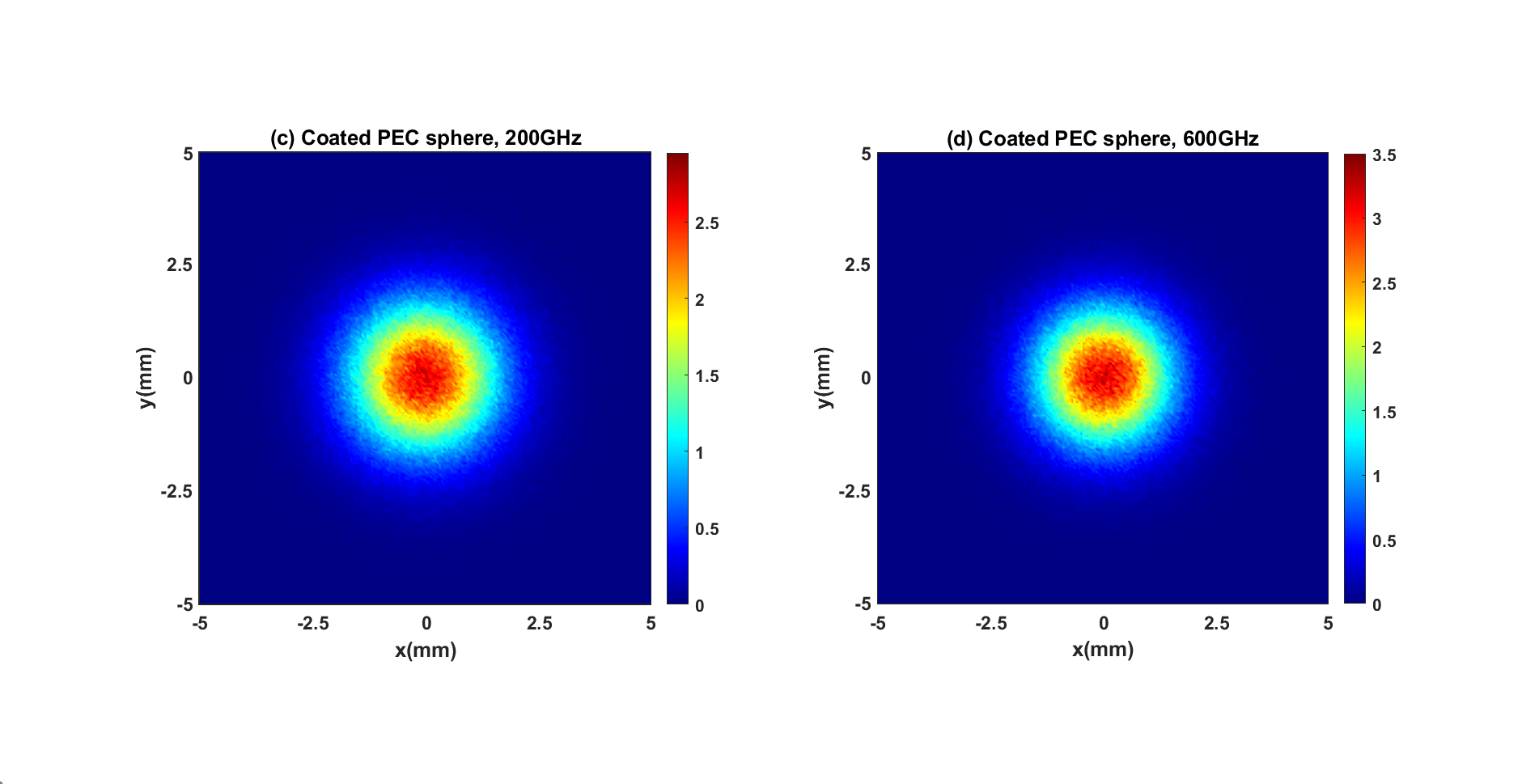}
\caption{Incoherent irradiance for $S_1$ strategy in 200 GHz and 600 GHz}\label{MQ}
\end{figure}

The FO results suggest that the lossless cornea shell is acting as a lens and improving the wavefront match to the $7$ mm RoC inner PEC surface relative to the reference $7.5$ mm RoC PEC sphere.
Non-sequential ray-tracing simulations of a Gaussian beam in Zemax OpticStudio also suggest that the presence of a lossless dielectric layer on top of the PEC (hemisphere) sphere might slightly improve the coupling. In Fig. \ref{MQ}, the power distribution on the detector is reported for $S_1$ at two different frequencies, 200 GHz and 600 GHz. At 200 GHz the root means square (RMS) spot radius is 3.40 mm for the coated sphere and 3.33 mm for the PEC sphere. At 600 GHz the RMS spot radius variation is much more contained: 3.13 mm for the coated sphere and 3.12 mm for the PEC sphere. This trend agrees with Fig. \ref{Q}(a) where the coupling shows a peak at 200 GHz.

In a quasi-optical, mono-static ($S_{11}$) measurement, the maximum signal is obtained when the beam is retro-directive, i.e. the scattered beam mirrors the illumination beam. In the case of a spherical target, this can be visualized by discretizing the wavefront as a spatial collection of converging rays. The maximum signal is achieved when all converging rays are normal to the spherical surface. However, in this case, the converging beam is Gaussian with a non-spherical phase front thus normal incidence is achieved only on-axis. The loss-free shell assists the incoming beam to come closer to the optical axis, that's why the above one values for coupling efficiency appeared in Fig. \ref{Q}.
\begin{table}[ht]
\centering
\caption{Number of modes for coated air sphere simulations}\label{t2}
\begin{tabular}{ c | c  c  c  c  c  c }
\hline
\textbf{Frequency (GHz)} & 100-190  & 190-285 & 285-350 & 350-445 & 445-510 & 510-600 \\ 
\hline
$\mathbf{N_{stop}}$ & 80 & 110 & 130 & 155 & 175 & 205 \\ [0.5ex] 
\hline
\end{tabular}
\end{table}

To carry on, a lossless cornea shell sitting on an air-core was simulated. The algorithm addressed in appendix B was used for calculating scattering coefficients of a dielectric coated sphere under plane wave illumination. Compared to other targets, simulations for coated air spheres needed more modes to reach stability due to the existence of multiple reflection beams resembling spherical cavities. Table \ref{t2} gives the number of modes used for simulation according to frequency. The simulation time for this target was about 75 hours on UPC computational cluster.

\begin{figure}[t]
\centering
\includegraphics[trim=90 125 90 145, clip,width=\textwidth]{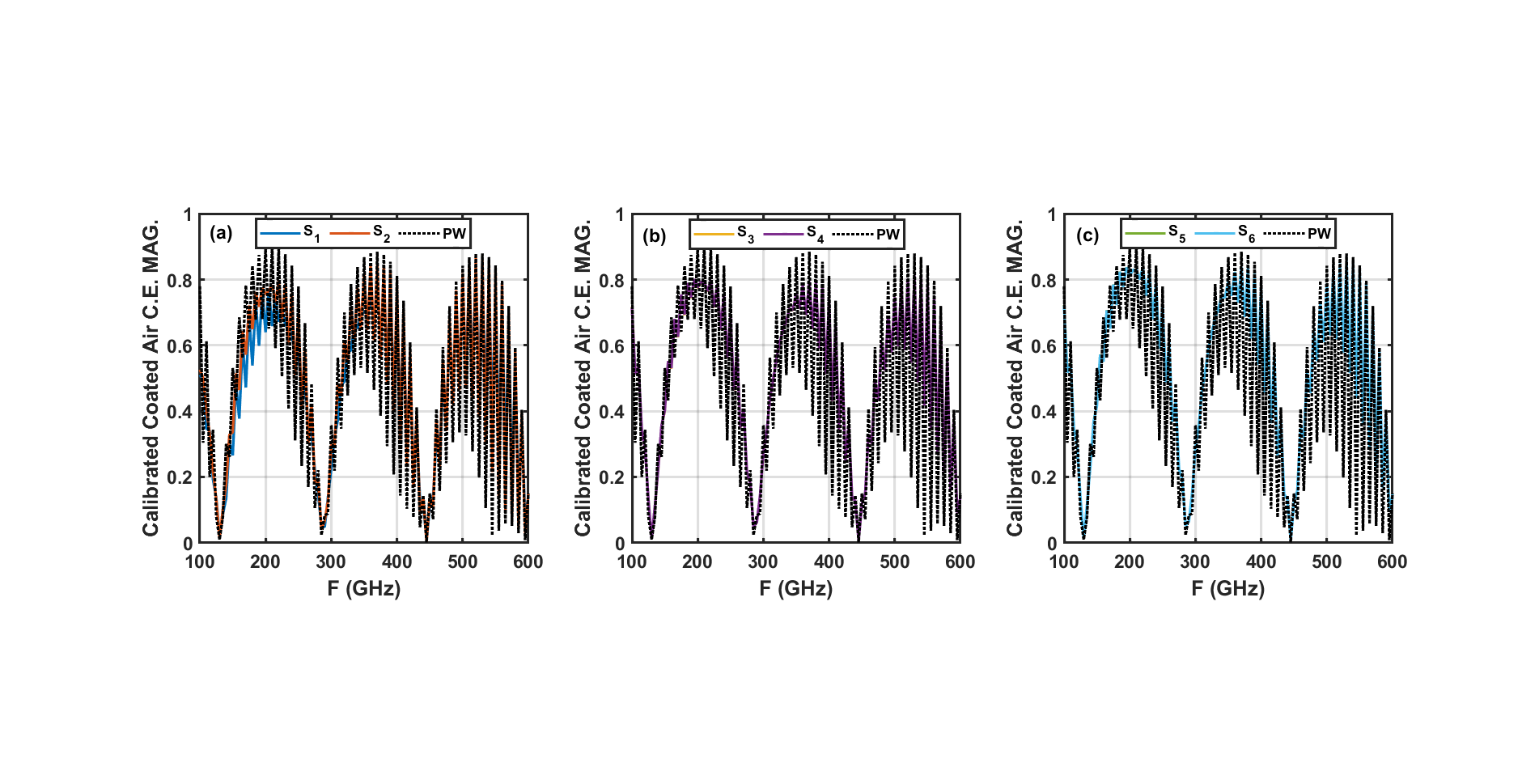}
\vspace*{-6 mm}

\includegraphics[trim=90 125 90 145, clip,width=\textwidth]{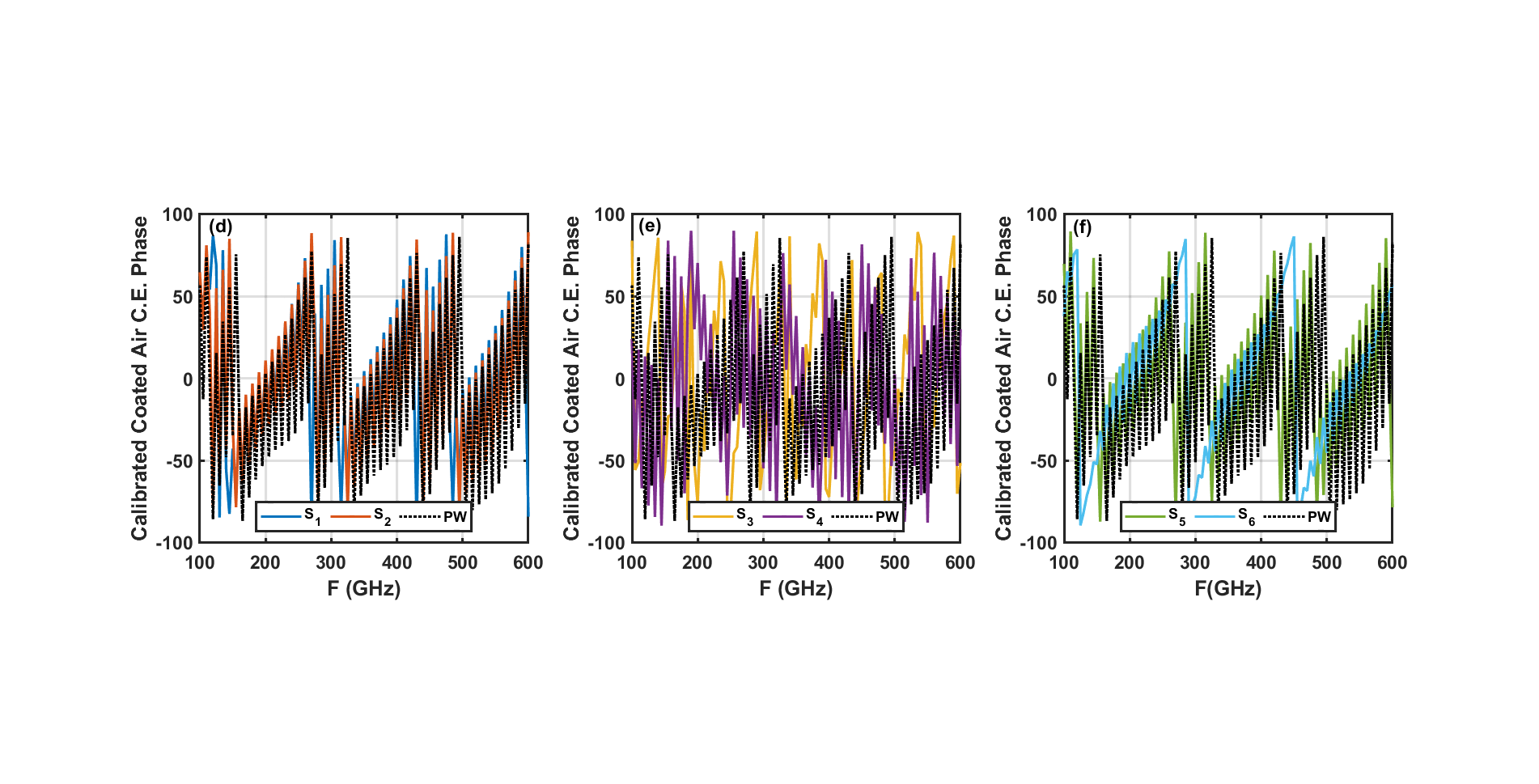}
\caption{Calibrated coated air sphere coupling efficiency magnitude and phase comparison are presented for (a,d) Forward strategies, $S_1$, and $S_2$, (b,e) reverse strategies, $S_3$, and $S_4$, and (c,f) reference strategies, $S_5$, and $S_6$. Plane wave reflection from equivalent layered planar is indicated by black-dotted line. The permittivity is set according to real part of cornea permittivity and its thickness is $0.5$ mm sitting on $7$ mm RoC air sphere.}\label{air}
\end{figure}

The calibrated magnitude and phase coupling efficiency for this target is plotted for all strategies in Fig. \ref{air}. The $S_2$ strategy is almost coherent with the equivalent planar structure for both phase and magnitude, indicating excellent coupling between the incident and back-scattered beam, especially for higher frequencies. From the magnitude aspect, reference strategies act the same and show higher coupling rather than reverse ones (which also behave the same), and eventually, the $S_1$ strategy owns the least consistency with plane wave condition. Phase coupling efficiency order from high to low is as follow $S_2>S_5>S_1>S_6>S3=S4$, emphasizing the lowest phase match in reverse strategies. However, the $S_4$ trend is more likely to plane wave condition only with $180^o$ phase shift.

\subsection{Calibrated cornea coupling efficiency}

The approach described in \cite{eucap}, and outlined in Appendix B, yielded the magnitude and phase coupling between the incidence and back-scattered electric fields for cornea structure. The number of modes used for simulation is reported in table \ref{t3} and simulation time was 25 hours on UPC computational cluster. As mentioned earlier, the best strategy for performing an experiment that matches with the theory developed in previous works \cite{quasi} is the one that Gaussian-beam illumination on a spherical surface acts such as a plane-wave illumination on a plane resembling a maximum coupling. 
\begin{table}[ht]
\centering
\caption{Number of modes for cornea simulations}\label{t3}
\begin{tabular}{ c | c  c  c  c }
\hline
\textbf{Frequency (GHz)} & 100-200 & 200-380 & 380-510 & 510-600 \\ 
\hline
$\mathbf{N_{stop}}$ & 65 & 75 & 100 & 125 \\ [0.5ex] 
\hline
\end{tabular}
\end{table}

\begin{figure}[ht]
\centering
\includegraphics[trim=90 125 90 125, clip,width=\textwidth]{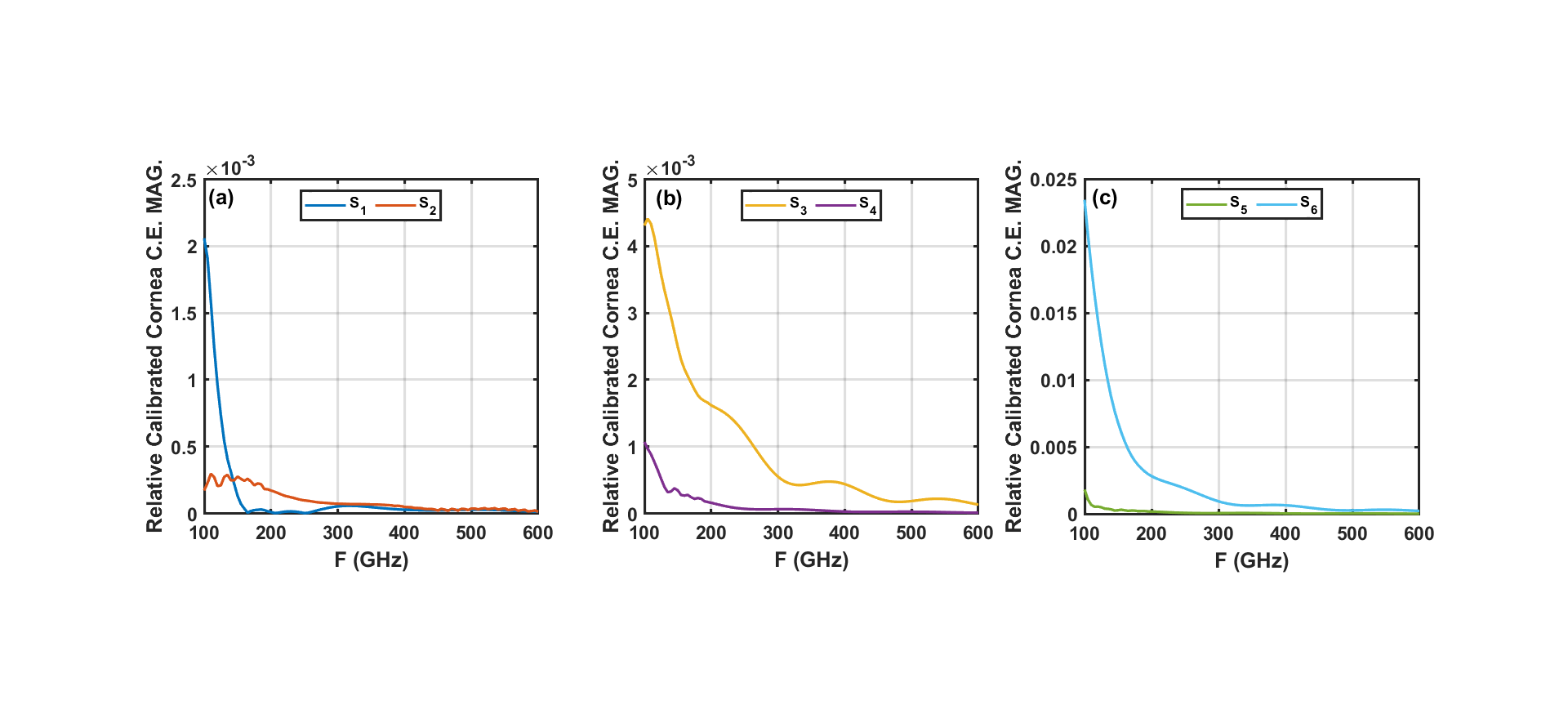}
\vspace*{-6 mm}

\includegraphics[trim=90 125 90 125, clip,width=\textwidth]{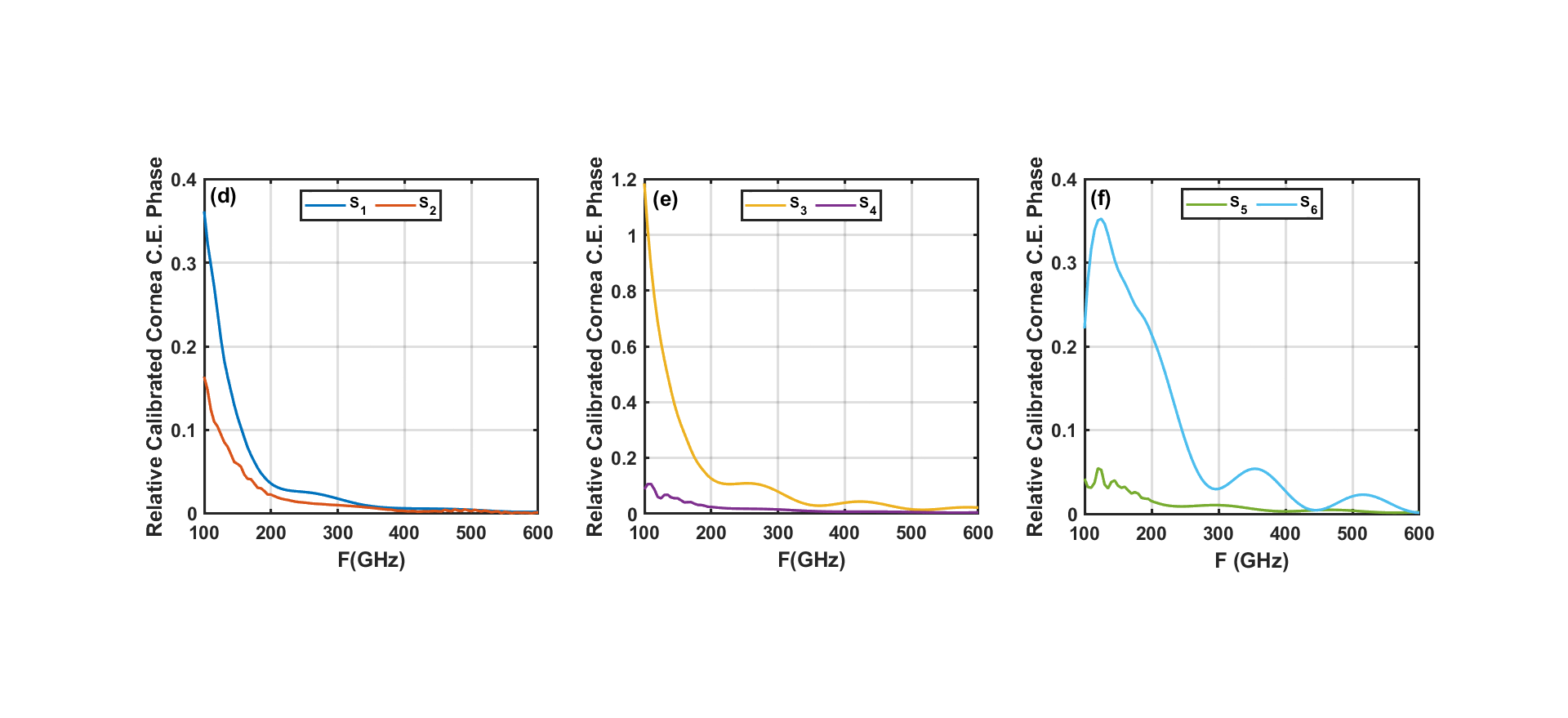}
\caption{Relative calibrated cornea coupling efficiency magnitude and phase comparison are shown for (a,d) Forward strategies, $S_1$, and $S_2$, (b,e) reverse strategies, $S_3$, and $S_4$, and (c,f) reference strategies, $S_5$, and $S_6$.}\label{cornea}
\end{figure}
 
For further exploration of different strategies for cornea target, relative corneal calibrated coupling efficiency magnitude and phase for each scheme were computed in Fig. \ref{cornea}, in this context, relative calibrated means after calibrating cornea coupling efficiency with equal RoC PEC sphere, it was subtracted from plane wave condition coupling efficiency.

Calibration with PEC sphere mirrors current experimental methodology where the reflectivity of a phantom or the ex-vivo cornea is normalized by a spherical reference reflector of approximately equal RoC. The reflectivity is obtained on an absolute scale when the optical path is deconvolved from the reflection by the normalization. These results support the previous hypothesis in the literature that, under the right alignment conditions, Gaussian-beam illumination on cornea approximates plane-wave illumination on planar stratified media.

 Surprisingly, these strategies behave almost the same as a plane-wave illumination on the planar surface and show a maximum of $2.3 \%$ magnitude deviation in low frequency for strategy $S_6$. Similarly, for calibrated phase coupling efficiency (indicated in Fig. \ref{cornea}(d-f)) of preceding strategies attributes almost the same as plane-wave illumination on the plane, with limited deviation at low frequency for strategy $S_3$ (a maximum of $1.2^{\circ}$ deviation).

These results suggest that, under ideal alignment and calibration conditions, the efficiency of coupling to corneal longitudinal modes is nearly indistinguishable amongst the six different strategies. This behavior is almost similar to the loss-free shell results, although the lossless cornea shell deviates from plane wave condition more than cornea case, indicating that lossy cornea steady state reflectivity minimizes walk-off losses and any illumination condition, regardless of wavefront match/mismatch is sufficient for proper coupling as far as equal surface geometry is feasible for calibration. 

\subsection{Sensitivity of each illumination strategy to PEC RoC error }

\begin{figure}[ht]
\centering
\includegraphics[trim=90 120 90 120, clip,width=\textwidth]{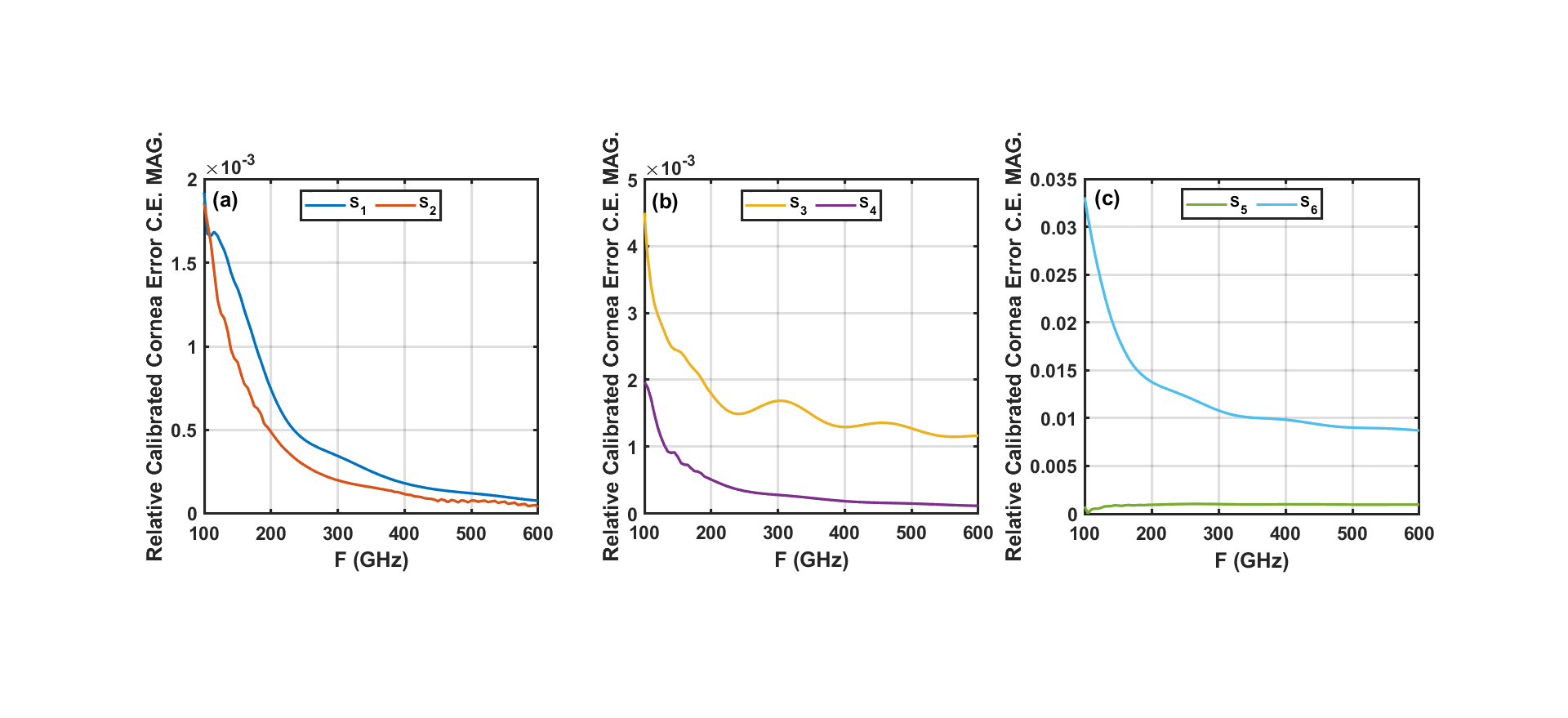}
\vspace*{-6 mm}

\includegraphics[trim=90 120 90 120, clip,width=\textwidth]{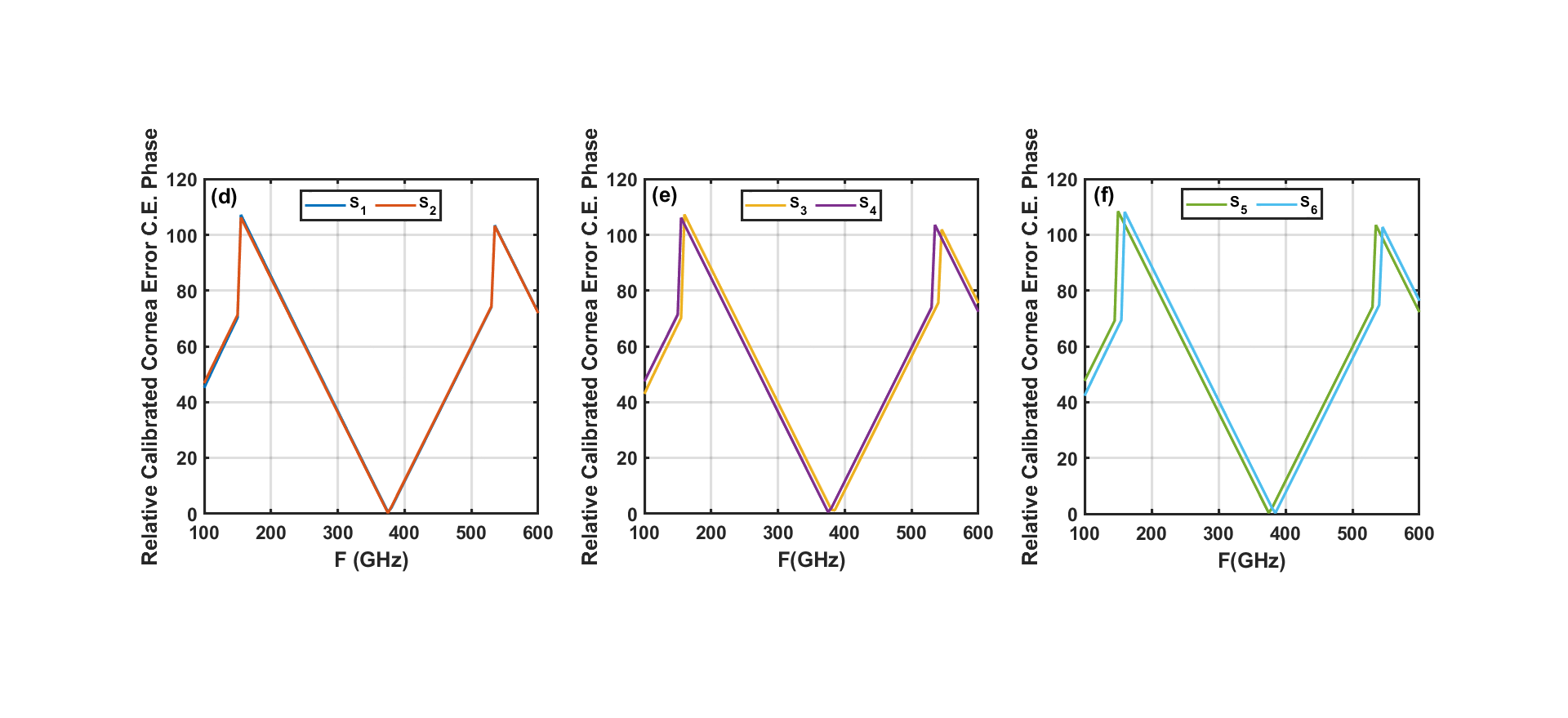}
\caption{ Relative calibrated cornea coupling efficiency magnitude and phase comparison when calibrated by $7.7$ mm radius PEC sphere are displayed for (a,d) Forward strategies, $S_1$, and $S_2$, (b,e) reverse strategies, $S_3$, and $S_4$, and (c,f) reference strategies, $S_5$, and $S_6$.}\label{error}
\end{figure}

 As mentioned, the PEC sphere was attended as a calibration target for simulations. To explore the sensitivity of each illumination strategy to the calibration error, the calibrated corneal coupling efficiency was recomputed using a $7.7$ mm RoC PEC sphere for calibration. Here, the coupling efficiency magnitude of the $7.7$ mm RoC PEC sphere was computed, aiming to show how the calibration is robust in case of discrepancies in the PEC sphere RoC. The required number of modes is similar to table \ref{t1}. Fig. \ref{error} shows the relative calibrated coupling efficiency magnitude and phase of a cornea when in PEC sphere existed $+0.2$ mm error in the RoC.

 The calibrated coupling efficiency of $S_3$ and $S_6$ are the most sensitive to RoC variation. The $S_6$ shows the largest deviation, $\sim 3.3 \%$. All other beam configurations deviate, at maximum, less than $\sim 0.5 \%$. The calibrated error analysis confirms that forward strategies are the most robust ($S_1$, $S_2$) strategies among the others and $S_4$ acts in a like manner and $S_5$ behaves closely to $S_3$. The phase plots in Fig. \ref{error}(d-f) reveal significant deviation for all strategies. Therefore, phase matching suffers dramatically from errors in calibration.

\section{Conclusion} 
A theory based on Fourier-optics analysis and VSH was employed to explore the coupling efficiency between the incident and back-scattered fields from homogeneous and coated spherical targets at different beam-target alignments. In this method, Gaussian beam approximated by an ensemble of plane waves and scattering coefficients were calculated via the T-matrix method. 

PEC spheres are targets for calibration in different simulations. PEC coupling efficiency magnitude in six different strategies was explored. The $S_2$ alignment shows the highest matching with plane-wave illumination on a planar surface. Also, the influence of $+0.2$-mm RoC change was investigated. The $S_6$ strategy shows the highest sensitivity (close to $3.3\%$) to RoC discrepancies. 

To explore the effect of spectral phase front variation and mismatch, coated PEC structure was studied. Due to the lossless layer in this assembly, walk-off loss takes part in the coupling, and beam profile influences the calibrated coupling efficiency magnitude, although phase behavior remains insensitive to the illumination profile and only shows $10$ GHz shift with respect to low frequencies than the plane wave condition. The $S_2$ design revealed the highest consistency with plane wave on plane illumination.

Surprisingly, the calibrated coupling efficiency magnitude for $S_1$ - $S_4$ strategies (in coated PEC) is above one. The reason is the refraction of the Gaussian beam while passing through the loss-free layer, as an incident beam propagates closer to the optical axis, the beam reflection enhances for certain kinds of Gaussian beams. This is in striking contrast to the stratified medium theory and theories considering plane waves - the specifics of the Gaussian beam and target RoC need to be considered carefully before applying the stratified medium model.

Also, a lossless cornea shell sitting on an air-core was considered as a target. In contrast with PEC and coated PEC targets, the $S_2$ strategy showed better matching rather than $S_1$, both for phase and magnitude. Reference strategy behaved better than reverse ones from the magnitude aspect. Overall, from a phase point of view, reverse strategies revealed the least consistency with plane wave conditions.

We used the Fourier analysis to investigate the corneal coupling efficiency. Cornea modeled as a layered media on top of the aqueous core which is illuminated by a Gaussian beam. It was shown that engineering the sub-millimeter wave beam such that incident beam $1/e$ radius fixed to $\omega_1=3.1$ mm and radius of the Gaussian beam matches with $7.5$ mm cornea radius at each frequency (strategy $S_2$) leads to the highest coupling among other strategies. 

Illumination profile defines the coupling efficiency behavior unless we calibrate the cornea coupling efficiency by a PEC sphere of the same size. In all strategies, coupling efficiency magnitude and phase behave according to the plane-wave illumination on the planar half-space. It implies that the cornea is lossy enough that walk-off loss does not influence coupling efficiency. 

The calibrated corneal coupling efficiency magnitude behaved like a plane wave illumination on the plane even with the presence of error in the calibrating PEC sphere although $S_6$ is less reliable. The phase behavior was dramatically sensitive to error and highly deviates from the phase of plane wave illumination on the plane.  

\begin{appendices}

\section{Gaussian beam coefficients derived by Fourier analysis}

For a polarized Gaussian beam in $x$ and $z$ direction, after a cumbersome calculation, the incident beam coefficients in (\ref{eq3}) are obtained as:

\begin{equation}\label{eq1_a}
\begin{gathered}
a_{e}=F\sum_{\theta_x}\sum_{\theta_y}4i^na^e_{\theta{xy}},\qquad
a_{o}=F\sum_{\theta_x}\sum_{\theta_y}4i^na^o_{\theta{xy}},\\
b_{e}=F\sum_{\theta_x}\sum_{\theta_y}-4i^{n+1}b^o_{e\theta{xy}},\qquad
b_{o}=F\sum_{\theta_x}\sum_{\theta_y}-4i^{n+1}b^o_{\theta{xy}},\\
\end{gathered}
\end{equation}
where $F=(k\omega_0p)^2/4\pi$ in which $p=\pi/180$. The $\theta_x$ and $\theta_y$ are angles respect to $x$ and $y$ axis. Also, 

\begin{equation}\label{eq2_a}
\begin{gathered}
 a^e_{\theta_{xy}}=A_{\theta_{xy}}(\sin\phi\cos m\phi\tau-m\cos\phi\sin m\phi\Pi/\cos\theta),\\
 a^o_{\theta_{xy}}=A_{\theta_{xy}}(m\cos\phi\cos m\phi\Pi/\cos\theta+\sin\phi\sin m\phi\tau),\\
 b^e_{\theta_{xy}}=A_{\theta_{xy}}(m\sin\phi\sin m\phi\Pi+\cos\phi\cos m\phi\tau/\cos\theta),\\
 b^o_{\theta_{xy}}=A_{\theta_{xy}}(\cos\phi\sin m\phi\tau/\cos\theta-m\sin\phi\cos m\phi\Pi).
\end{gathered}
\end{equation}

 In the above equations, the $\Pi=\Pi_{mn}=\frac{P^m_n(\cos\theta)}{\sin\theta}$ and $\tau=\tau_{mn}=\frac{d}{d\theta}P^m_n(\cos\theta)$ are auxiliary functions which can be obtained by recursion relation \cite{barber}: 
 
\begin{equation}\label{eq3_a}
\begin{gathered}
\Pi_{mn}=\frac{(2n-1)\cos\theta\Pi_{mn-1}-(n+m-1)\Pi_{mn-2}}{n-m},\\
\tau_{mn}=n\cos\theta\Pi_{mn}-(n+m)\Pi_{mn-1}.
\end{gathered}
\end{equation}
  
 The first two starting values of $\Pi_{mn}$ are generated by the closed-form expressions when $m\neq0$:
 
\begin{equation}\label{eq4_a}
\begin{gathered}
 \Pi_{mn}=0\qquad n<m,\\ \Pi_{mn}=\frac{(2m)!\sin^{m-1}(\theta)}{2^mm!}\qquad n=m,
\end{gathered}
\end{equation}
and while $m=0$:
 \begin{equation}\label{eq5_a}
 \begin{gathered}
 \Pi_{0,0}=\frac{1}{\sin\theta},\qquad
 \Pi_{0,1}=\frac{\cos\theta}{\sin\theta}.\\
 \end{gathered}
 \end{equation}

On the other hand, the $A_{\theta_{xy}}=TG\sin\theta_x\sin\theta_y$ where $T=e^{-ik(\textbf{s}.\textbf{v})}=e^{-i(k_xx_0+k_yy_0+k_zz_0)}$ indicates the relocation of waist radius location relative to center of sphere and $G=e^{-(\frac{ks\omega_0}{2})^2}=e^{-\frac{\omega^2_0}{4}(k^2_x+k^2_y)}$. Besides, $PW=e^{ik(\textbf{s}.\textbf{r})}=e^{i(k_xx+k_yy+k_zz)}$ illustrates a plane wave. The vectors $\textbf{s}$, $\textbf{v}$, and $\textbf{r}$ are $\textbf{s}=\cos\theta_x\mathbf{i_x}+\cos\theta_y\mathbf{i_y}+\cos\theta\mathbf{i_z}$ , $\textbf{v}=x_0\mathbf{i_x}+y_0\mathbf{i_y}+z_0\mathbf{i_z}$, and $\textbf{r}=x\mathbf{i_x}+y\mathbf{i_y}+z\mathbf{i_z}$. The wave vector defines as $\textbf{k}=k_x\mathbf{i_x}+k_y\mathbf{i_y}+k_z\mathbf{i_z}$ and $k=|\textbf{k}|=[k_x^2+k_y^2+k_z^2]^{1/2}=2\pi/\lambda$ is the wave number in the wavelength $\lambda$. For any arbitrary $\textbf{k}$, $\theta_x$, $\theta_y$, $\theta$, and $\phi$ are defined the way that the components of $\textbf{k}$ are:
\begin{equation}\label{eq6_a}
\begin{gathered}
k_x=k\cos\theta_x=k\sin\theta\cos\phi,\\
k_y=k\cos\theta_y=k\sin\theta\sin\phi,\\
k_z=k\cos\theta.
\end{gathered}
\end{equation}

\section{Coated sphere illuminated by a plane wave}

The scattering coefficients of a coated sphere illuminated by a plane wave is presented by \cite{Yang} and summarized in \cite{pena}. The method starts with solving the Helmholtz equation in spherical coordinate and fulfillment of the boundary condition at each layer. The scattering coefficients are derived as below:

\begin{equation}\label{eq1_b}
\begin{gathered}
a^c_n=\frac{[H^a_n(m_lx_l)/m_l+n/x_l]\psi_n(x_l)-\psi_{n-1}(x_l)}{[H^a_n(m_lx_l)/m_l+n/x_l]\zeta_n(x_l)-\zeta_{n-1}(x_l)},\\
b^c_n=\frac{[H^b_n(m_lx_l)/m_l+n/x_l]\psi_n(x_l)-\psi_{n-1}(x_l)}{[H^b_n(m_lx_l)/m_l+n/x_l]\zeta_n(x_l)-\zeta_{n-1}(x_l)},
\end{gathered}
\end{equation}
where $\psi_n$ and $\zeta_n$ are Riccati-Bessel functions. The $m_l$ and $x_l$ are refractive index and size parameter of the $l$th layer. The $H^a_n$ and $H^a_n$ are given by following expressions:

\begin{equation}\label{eq2_b}
\begin{gathered}
H^a_n(m_1x_1)=D^1_n(m_1x_1),\\
H^a_n(m_lx_l)=\frac{G_2D^1_n(m_lx_l)-Q^l_nG_1D^3_n(m_lx_l)}{G_2-Q^l_nG_1},\\
H^b_n(m_1x_1)=D^1_n(m_1x_1),\\
H^a_n(m_lx_l)=\frac{\hat G_2D^1_n(m_lx_l)-Q^l_n\hat G_1D^3_n(m_lx_l)}{\hat G_2-Q^l_n\hat G_1},\\
G_1=m_lH^a_n(m_{l-1}x_{l-1})-m_{l-1}D^1_n(m_lx_{l-1}),\\
G_2=m_lH^a_n(m_{l-1}x_{l-1})-m_{l-1}D^3_n(m_lx_{l-1}),\\
\hat G_1=m_{l-1}H^b_n(m_{l-1}x_{l-1})-m_lD^1_n(m_lx_{l-1}),\\
\hat G_1=m_{l-1}H^b_n(m_{l-1}x_{l-1})-m_lD^3_n(m_lx_{l-1}).\\
\end{gathered}
\end{equation}

Next equations determine the logarithmic derivatives of the Riccati–Bessel
functions, $D^1_n(z)$ and $D^3_n(z)$, the ratio $Q^l_n$, $\psi_n$ and $\zeta_n$. A thorough explanation can find in \cite{Yang} and \cite{pena}.

\begin{equation}\label{eq3_b}
\begin{gathered}
D^1_{N_{max}}(z)=0+0i,\hspace{1em plus0.5em}
D^1_{n-1}(z)=\frac{n}{z}-\frac{1}{D^1_{n}(z)+\frac{n}{z}},\\
D^3_0(z)=i,\hspace{1em plus0.5em} D^3_{n}(z)=D^1_{n}(z)+\frac{i}{\psi_n(z)\zeta_n(z)},\\
\psi_0(x_l)\zeta_0(x_l)=\frac{1}{2}[1-(\cos 2a+i\sin 2a)\exp(-2b)],\\
\psi_n(x_l)\zeta_n(x_l)=\psi_{n-1}(x_l)\zeta_{n-1}(x_l)[\frac{n}{z}-D^1_{n-1}(z)][\frac{n}{z}-D^3_{n-1}(z)],\\
\psi_0(x_l)=\sin(x_l),\hspace{1em plus0.5em}\psi_n(x_l)=\psi_{n-1}(x_l)[\frac{n}{x_l}-D^1_{n-1}(x_l)],\\
\zeta_0(x_l)=\sin(x_l)-i\cos(x_l),\hspace{1em plus0.5em}\zeta_n(x_l)=\zeta_{n-1}(x_l)[\frac{n}{x_l}-D^3_{n-1}(x_l)],\\
Q^l_0=\frac{\exp(-2ia_1)-\exp(-2b_1)}{\exp(-2ia_2)-\exp(-2b_2)}\exp(-2[b_2-b_1]),\\
Q^l_n=Q^l_{n-1}(\frac{x_{l-1}}{x_l})^2\frac{[z_2D^1_n(z_2)+n]}{[z_1D^1_n(z_1)+n]}\frac{[n-z_2D^3_{n-1}(z_2)]}{[n-z_1D^3_{n-1}(z_1)]},
\end{gathered}
\end{equation}
where $z=a+ib$, $z_1=m_lx_{l-1}=a_1+ib_1$ and $z_2=m_lx_l=a_2+ib_2$. For all recurrence relation $n=1,..,N_{max}$ except $D^1_n$ which is a downward recurrence. The maximum number of the modes $N_{max}$ play an crucial rule to stability of the problem. It is a function of the size parameter and $N_{max}=max(N_{stop},|m_lx_l|,|m_lx_{l-1}|)+15$ when $l=1,2,..,L$ and 

\begin{equation}\label{eq30}
N_{stop}=\begin{cases}x_l+4x_l^{1/3}+1 & 0.02 \leq x_l< 8\\
x_l+4.05x_l^{1/3}+2 & 8\leq x_l< 4200\\
x_l+4x_l^{1/3}+2 & 4200\leq x_l< 20,000.\\
\end{cases}
\end{equation}

As mentioned in section 3, to compute the scattering coefficients of a coated sphere while illuminated by a plane wave, a different method compared to \cite{esam93} and \cite{esam94} is applied. Khaled used Toon and Ackerman \cite{Toon} algorithm while we used the algorithm described by Yang \cite{Yang}. The advantage of the Yang method is providing the possibility of analysis of a coated sphere with more than one shell.

\section{PEC and coated PEC sphere illuminated by a plane wave}

The scattering coefficients of a sphere illuminated by a plane wave are computed by solving the vector-wave equation in spherical coordinates and evoking the boundary condition (Mie theory \cite{bohrn}). In the case of a PEC sphere, we force the electric field zero inside the PEC boundary and the simplified form of scattering coefficients for a PEC sphere are obtained as:

\begin{equation}\label{eqc1}
\begin{array}{l}
a_n=\frac{\psi_n(x)'}{\zeta_n(x)'}\\
b_n=\frac{\psi_n(x)}{\zeta_n(x)},
\end{array}
\end{equation} 
where the size parameter $x=kr$ and $r$ is the radius of the PEC sphere.

With the mutual approach, solving a vector-wave equation, applying boundary condition and forcing the electric field to be zero inside the PEC core, the simplified scattering coefficients for a coated PEC sphere are obtained as: 

\begin{equation}\label{eqc2}
\begin{array}{l}
a_n=\frac{\psi_n(x_1)[\psi_n'(m_2x_1)-A_n\chi_n'(m_2x_1)]-m_2\psi_n'(x_1)[\psi_n(m_2x_1)-A_n\chi_n(m_2x_1)]}{\zeta_n(x_1)[\psi_n'(m_2x_1)-A_n\chi_n'(m_2x_1)]-m_2\zeta_n'(x_1)[\psi_n(m_2x_1)-A_n\chi_n(m_2x_1)]},\\

b_n=\frac{m_2\psi_n(x_1)[\psi_n'(m_2x_1)-B_n\chi_n'(m_2x_1)]-\psi_n'(x_1)[\psi_n(m_2x_1)-B_n\chi_n(m_2x_1)]}{m_2\zeta_n(x_1)[\psi_n'(m_2x_1)-B_n\chi_n'(m_2x_1)]-\zeta_n'(x_1)[\psi_n(m_2x_1)-B_n\chi_n(m_2x_1)]},\\
\end{array}
\end{equation} 
where 
\begin{equation}\label{eqc3}
\begin{array}{l}
A_n=\frac{\psi_n(m_2x)'}{\chi_n(m_2x)',} \\ 
B_n=\frac{\psi_n(m_2x)}{\chi_n(m_2x)}.
\end{array}
\end{equation} 
The PEC core radius $r_1$ is coated with a layer with outer radius of $r_2$ and refractive index of $m_2$. The Riccati-Bessel functions are $\chi_n(z)=-zy_n(z)$, $\psi_n(z)=zj_n(z)$, and $\zeta_n(z)=zh_n^{(1)}(z)$ in which $j_n(z)$, $y_n(z)$, and $h_n^{(1)}(z)$ are Bessel function of the first, Bessel function of the second kind and Hankel function of the first kind, respectively. The primes denote the differentiation with respect $z$.
\end{appendices}

\section*{Funding}
This research was supported by the Agencia Estatal Investigación PID2019-107885GB-C31/AEI/10.13039,
PRE2018-084326, and Academy of Finland Project number 327640.

\section*{Acknowledgments}
We would like to thank Prof. Ari Sihvola for his valuable advice to understand and implement the theory of problem.

\section*{Disclosures}
The authors declare that there are no conflicts of interest related to this article.











\end{document}